\documentclass[a4paper,twocolumn, numberedappendix,revtex4,iop]{oja} 
\usepackage[a4paper,marginpar=10pt,headheight=20pt, right=1cm, top=1.5in, left=1in, bottom=0pt, footskip=0pt, footnotesep=10pt]{geometry}
\usepackage[T1]{fontenc}
\usepackage[utf8]{inputenc}
\usepackage{natbib}
\usepackage{multirow}
\usepackage{graphicx} 
\usepackage{amsmath}	
\usepackage{amssymb}	
\usepackage{longtable}  
\usepackage{rotating}  
\usepackage{xcolor}          
\usepackage{booktabs}
\usepackage{xurl}
\usepackage{microtype}
\usepackage[colorlinks,linkcolor=blue,citecolor=blue,urlcolor=blue ]{hyperref}

\DeclareRobustCommand{\sbseries}{\fontseries{sb}\selectfont}
\DeclareTextFontCommand{\textsb}{\sbseries}

\DeclareRobustCommand{\sbseries}{}

\newcommand{\Tagn}{\log_{10}(\frac{T_\mathrm{AGN}}{\mathrm{K}})}
\newcommand{\ks}{\log_{10}(\frac{k_\mathrm{s}}{\mathrm{Mpc}^{-1}})}
\newcommand{\aB}{\hat{\alpha}_\mathrm{B}}
\newcommand{\aM}{\hat{\alpha}_\mathrm{M}}
\DeclareMathAlphabet{\baz}{OT1}{}{}{}
\begin{document}
\title{Fast Radio Bursts as a probe of gravity on cosmological scales}
\author{Dennis Neumann}
\affiliation{Leiden Observatory, Leiden University, P.O. Box 9513, 2300 RA Leiden, the Netherlands
}
\affiliation{Ruhr University Bochum, Faculty of Physics and Astronomy, Astronomical Institute (AIRUB), German Centre for Cosmological Lensing, 44780 Bochum, Germany}
\thanks{$^*$\href{mailto:dneumann@strw.leidenuniv.nl}{dneumann@strw.leidenuniv.nl}}

\author{Robert Reischke}
\affiliation{Argelander-Institut für Astronomie, Universität Bonn, Auf dem Hügel 71, D-53121 Bonn, Germany}
\affiliation{Ruhr University Bochum, Faculty of Physics and Astronomy, Astronomical Institute (AIRUB), German Centre for Cosmological Lensing, 44780 Bochum, Germany}
\thanks{$^\dagger$\href{mailto:reischke@posteo.net}{reischke@posteo.net}}

\author{Steffen Hagstotz}
\affiliation{Universitäts-Sternwarte, Fakultät für Physik, Ludwig-Maximilians Universität München, 
Scheinerstraße 1, D-81679 München, Germany and\\
Excellence Cluster ORIGINS, Boltzmannstraße 2, D-85748 Garching, Germany}

\author{Hendrik Hildebrandt}
\affiliation{Ruhr University Bochum, Faculty of Physics and Astronomy, Astronomical Institute (AIRUB), German Centre for Cosmological Lensing, 44780 Bochum, Germany}

\begin{abstract}
We explore the potential for improving constraints on gravity by leveraging correlations in the dispersion measure derived from Fast Radio Bursts (FRBs) in combination with cosmic shear. Specifically, we focus on Horndeski gravity, inferring the kinetic braiding and Planck mass run rate from a stage-4 cosmic shear mock survey alongside a survey comprising $10^4$ FRBs. For the inference pipeline, we utilise the {Boltzmann code} \texttt{hi\_class} to predict the linear matter power spectrum in modified gravity scenarios, while non-linear corrections {are obtained from the halo-model employed in \texttt{HMcode}}, including feedback mechanisms. Our findings indicate that FRBs can disentangle degeneracies between baryonic feedback and cosmological parameters, as well as the mass of massive neutrinos. Since these parameters are also degenerate with modified gravity parameters, the inclusion of FRBs can enhance constraints on Horndeski parameters by up to $40$ percent, despite being a less significant measurement. Additionally, we apply our model to current FRB data and use the uncertainty in the $\mathrm{DM}-z$ relation to impose limits on gravity. However, due to the limited sample size of current data, constraints are predominantly influenced by theoretical priors. Despite this, our study demonstrates that FRBs will significantly augment the limited set of cosmological probes available, playing a critical role in providing alternative tests of feedback, cosmology, and gravity. All codes used in this work are made publicly available.
\end{abstract}

\maketitle
\section{Introduction}
The Universe entered a phase of accelerated expansion around four billion years ago \citep{riess_observational_1998}. This effect is one of the greatest unsolved mysteries of modern cosmology and understanding the mechanisms behind this \emph{dark energy} is the proclaimed science goal of the stage-4 cosmological surveys Euclid\footnote{\href{https://www.euclid-ec.org/}{https://www.euclid-ec.org/}}, Rubin-LSST\footnote{\href{https://www.lsst.org/}{https://www.lsst.org/}} and the Roman telescope\footnote{\href{https://roman.gsfc.nasa.gov/}{https://roman.gsfc.nasa.gov/}}. In the standard model of cosmology, dark energy is viewed as a perfect fluid driven by the cosmological constant $\Lambda$. Even though this picture is well motivated at the moment \citep{Tripathi_2017_dark,planck_collaboration_planck_2020,Escamilla_2024_state}, accurate modelling of the dark energy dynamics is of utmost importance due to its implication of large-scale structure (LSS) formation. The recent DESI findings of a possibly time-dependent dark energy equation of state \citep{DESI_2024_cosmological} further compel us to explore alternative gravity models. The simplest modifications to GR involve a dynamic dark energy fluid driven by an additional scalar degree of freedom in the relativistic field equations. A plethora of these modifications are being actively investigated on the cosmological level, including but not limited to \emph{Quintessence} \citep{ratra_expressions_1988}, \emph{Chameleon} \citep{khoury_2004_chameleon} and $f(R)$ \citep{carroll_2004_cosmic}. Such theories, including GR, can be generalised under \emph{Horndeski theory} \citep{horndeski_second-order_1974}; the most general scalar-tensor theory that has second-order field equations \citep{kobayashi_horndeski_2019}. \\
\indent Even though GR is an excellent description of gravity on small cosmic scales, small deviations on larger scales have not been ruled out, yet \citep{alonso_2017_observational, Heisenberg_2018_large,reischke_investigating_2019,spurio_mancini_3d_2018, spurio_mancini_kids_2019}. Furthermore, differences in LSS between individual theories can be marginal, necessitating the development of precise cosmological tools. The aforementioned stage-4 surveys will map billions of galaxies which will subsequently be used for inferring cosmic evolution by means of statistical weak gravitational lensing (WL) analyses, amongst others \citep[see e.g.][]{weinberg_observational_2013}. 

While WL is a particularly strong probe for the underlying matter distribution and its ``clumpiness'' specifically \citep{secco_2022_DESY3}, current and future surveys are limited by uncertainties of baryonic processes \citep{tröster_2022_joint}. This problem will only become more severe with more resolved data and hence lower noise levels unveiling correlations on scales smaller than $1\mathrm{Mpc}$, {where baryonic feedback can suppress the power spectrum by up to $40\%$\,\citep{chisari_2019_modelling}}. In particular, the general shape of this suppression is unknown, thus, requiring tight priors in order to be used in cosmological inference.
A solution might be the steady rise of Fast Radio Burst (FRB) measurements \citep{lorimer_bright_2007,chime_2021_first}; broad, millisecond transient pulses in the radio frequency range that get dispersed by free electrons along their line of sight and that are consequently highly sensitive to baryons. Even though their origin is still debated \citep{petroff_fast_2019}, the majority of them must be of extragalactic origin due to their high Dispersion Measure (DM). Naturally, they have been proposed as a LSS probe \citep{zhou_fast_2014}, in particular since the Signal-to-Noise Ratio (SNR) of individual measurements is already ample \citep{yang_2016_extracting}. FRBs are versatile; they have been suggested as a tool to find the missing baryon content in the intergalactic medium \citep{mu_2018_finding}, or more generally as an $\Omega_\mathrm{b}h^2$ probe \citep{Walters_2018_future}. Furthermore, investigating modified gravity with them is not a new concept. Their low-resolution high-SNR nature was utilised in strong gravitational fields before\,(\citealt{adi_2021_probing}, \citealt{jiang_2024_exploring}) but the lack of host-identified FRBs prevents a statistically meaningful LSS analysis in the WL regime. The Square Kilometre Array (SKA\footnote{\href{https://www.skao.int/}{https://www.skao.int/}}, \citealt{dewdney_2009_square}) will address this problem at the latest \citep{hashimoto_2020_fast}. Until then, we need to identify the expected constraining power and, in particular, limitations of the currently available modelling tools. So far, investigations focused on how to leverage the averaged cosmological signal, the so-called $\mathrm{DM}-z$ relation \citep[e.g.][]{zhou_fast_2014,Walters_2018_future,hagstotz_new_2022,macquart_census_2020,wu_8_2022,james_measurement_2022,2023_reischke_consistent,reischke_cosmological_2023} or the statistics of DM fluctuations \citep[e.g.][]{masui_dispersion_2015,shirasaki_large-scale_2017,rafiei-ravandi_chimefrb_2021,bhattacharya_fast_2020,takahashi_statistical_2021,reischke_probing_2021, reischke_consistent_2022,reischke_calibrating_2023} to constrain the cosmological standard model.

In this work, we extend the previous work shown in \citet{reischke_calibrating_2023} and forecast the constraining power arising from a joint FRB+WL measurement on a broad selection of cosmological parameters, in particular those describing perturbations to GR in the form of Horndeski theory. We aim to quantify the improvement in precision when combining a virtual FRB survey feasible within the next decade with stage-4 data, as well as identify shortcomings of the currently available numerical modelling. In Section \ref{ch:theory} we first introduce Horndeski theory while focusing on its parameterisation implemented by the \texttt{hi\_class}\footnote{\href{http://miguelzuma.github.io/hi_class_public/}{http://miguelzuma.github.io/hi\_class\_public/}} \citep{zumalacarregui_hiclass:_2016} suite we are using. Afterwards, we lay out the angular power spectrum signal and noise for WL and FRB auto- and cross-correlation in modified gravity. Then, in Section \ref{ch:method}, we specify the mock catalogues used in this work. Afterwards, we explain how we obtain the likelihood based on emulated cosmological functions, namely the three-dimensional electron and matter power spectrum and the modified gravity change to the Poisson equation and line element in conformal Newtonian gauge. Here, we list the full set of fixed and inferred cosmological parameters with their respective fiducial value. With the emulated functions we obtain the likelihood and thus, through nested sampling with an MCMC-algorithm, the results displayed in Section \ref{ch:results}. There, we showcase the constraints for a small and a large deviation of gravitational dynamics from GR and explore the influence of our Horndeski theory parameter priors on the overall resulting contours. We argue that, while we see an improvement when cross-correlating FRBs with WL, we experience a loss in accuracy from underlying model limitations that must be considered when analysing real data as it becomes available. In Section \ref{ch:appli} we apply the outlined procedure to the covariance of FRBs with host identification.
A full summary of the results is found in Section \ref{ch:conclusion}.

The code, data and trained emulators used in this work are {publicly available via \texttt{git } on \href{https://github.com/DennisNeumann97/frb_horndeski_forecast}{https://github.com/DennisNeumann97/ frb\_horndeski\_forecast} for the forecast and on \href{https://github.com/rreischke/frb_covariance}{https://github.com/rreischke/frb\_covariance} for the LSS covariance of FRBs also described in \citet{reischke_cosmological_2023}.

\section{Theory}
\label{ch:theory}


\subsection{Horndeski theory}
\label{ch:theo_horndeski}

{As opposed to the standard cosmological model, the Lagrangian density in Horndeski modified gravity} depends on the metric tensor $g^{\mu\nu}$ \textit{and} an additional scalar degree of freedom $\phi$\,\citep{horndeski_second-order_1974}. It can be written as
\begin{align}
    \mathcal{L}[g_{\mu\nu},\phi,\psi] = \left(\sum_{n=2}^5 \mathcal{L}_n[g_{\mu\nu},\phi] + \mathcal{L}_\mathrm{M}[g_{\mu\nu},\psi]\right)
\end{align}
and enters the action according to 
\begin{align}
    \label{eq:theo_horndeski_action}
    S = \int \mathrm{d}^4x\sqrt{-g} \mathcal{L} \;.
\end{align}
$\mathcal{L}_\mathrm{M}[g^{\mu\nu}, \psi]$ denotes contribution of the matter field $\psi$ to the Lagrangian density. With the covariant derivative $\phi_{;\mu} := \nabla_\mu\phi$, the d'Alembert operator $\Box \phi := g^{\mu\nu} \phi_{;\mu\nu}$ and the scalar field canonical kinetic energy \linebreak $X := -g^{\mu\nu}\phi_{;\mu}\phi_{;\nu}/2$,  the $\mathcal{L}_n$ are given by
\begin{align}
    \mathcal{L}_2 & = G_2(\phi,\,X),\nonumber\\  
    \mathcal{L}_3 & = -G_3(\phi,X)\Box\phi, \nonumber\\ 
    \mathcal{L}_4 & = G_4(\phi,X)R+G_{4;X}(\phi,X)\left[(\Box\phi)^2-\phi^{;\mu\nu}\phi_{;\mu\nu}\right],  \nonumber\\ 
    \mathcal{L}_5 & = G_5(\phi,X) G^{\mu\nu}\phi_{;\mu\nu}\nonumber
     -\frac{1}{6}G_{5;X}(\phi,X)\\ & \times\left[(\Box\phi)^3- 3\Box\phi\phi^{;\mu\nu}\phi_{;\mu\nu}+2\phi_{;\mu}^{\phantom{;\mu}\nu}\phi_{;\nu}^{\phantom{;\nu}\lambda}\phi_{;\lambda}^{\phantom{;\lambda}\mu}\right]\;,
\end{align}
where $G_2$, $G_3$, $G_4$ and $G_5$ are arbitrary functions of $\phi$ and $X$. {\citet{bellini_maximal_2014} have redefined the $G_i$ into four time-dependent functions $\alpha_i$ that fully describe linear perturbations to the Horndeski action shown in Eq.\;\ref{eq:theo_horndeski_action}. With the help of the now time dependent Planck mass}
\begin{align}
    M_\mathrm{Pl}^2 := \;2\left(G_4+X(G_{5;\phi}-2G_{4;X}-H\dot{\phi}G_{5;X})\right) \nonumber
\end{align}
{they can be written as}:
\begin{align}
    H M_\mathrm{Pl}^2\,\alpha_\mathrm{M} := & \;\frac{\mathrm{d}}{\mathrm{d}t}M_\mathrm{Pl}^2\,  ,\nonumber  \\[0.5em]
    H M_\mathrm{Pl}^2\,\alpha_\mathrm{K} := & \; \nonumber\\
      2X (&G_{2;X}+2XG_{2;XX}-2G_{3;\phi}-2XG_{3;\phi X}) \nonumber\\
     +\,12\dot{\phi}XH (&G_{3;X}+X(G_{3;XX}-2G_{4;\phi XX}) - 3G_{4;\phi X})\nonumber\\
     \; +\, 12XH^2 (&G_{4;X}+8XG_{4;XX}+4X^2G_{4;XXX}) \nonumber\\
     \; +\,12XH^2 (&G_{5;\phi}+5XG_{5;\phi X}+2X^2G_{5;\phi XX})\nonumber\\
 \; +\,4\dot{\phi}XH^4 (&3G_{5;X}+7XG_{5;XX}+2X^2G_{5;XXX})\, , \nonumber \\[0.5em]
    H M_\mathrm{Pl}^2\,\alpha_\mathrm{B} := & \; 2\dot{\phi}\left(XG_{3;x}-G_{4;\phi}-2XG_{4;\phi X}\right) \nonumber \\
     +\;8XH\phantom{\phi^2}&\left(G_{4;X}+2XG_{4;XX}-G_{5;\phi}-XG_{5;\phi X}\right) \nonumber \\
     +\;2\dot{\phi}XH^2 &\left(3G_{5;X}+2XG_{5;XX} \right)\; ,\nonumber \\[0.5em]
    M_\mathrm{Pl}^2\,\alpha_\mathrm{T} := & \; 2X\left(2G_{4;X}-2G_{5;\phi} - \left(\ddot{\phi}-\dot{\phi}H\right) G_{5;X}\right)\,,
\end{align}
where $H$ denotes the time-dependent Hubble parameter. Physically, these functions can be interpreted as follows \citep{bellini_constraints_2016}:
\begin{itemize}
    \item \emph{Planck-mass run rate} $\alpha_\mathrm{M}$ \\
    In Horndeski-type theories, the Planck-Mass is not necessarily a constant. Its evolution through cosmic history is captured by $\alpha_\mathrm{M}$. A boundary condition is necessary to well-define the Planck mass. A sensible choice is to use the value it possesses today on small scales $M_\mathrm{Pl}^{\mathrm{ini}}$. This is a free parameter if a screening mechanism forces Horndeski theory to reduce to GR on small scales. Otherwise, it must be the observed value $M_\mathrm{Pl}^{\mathrm{ini}}=2.176 434(24)\cdot10^{-8}\,\mathrm{kg}$ \citep{tiesinga_codata_2021}. $\alpha_\mathrm{M}\neq0$ indicates a non-minimally coupled theory of gravity\footnote{Meaning the ``charge'' of gravity (the metric $g_{\mu\nu}$) has a more complicated relationship to the scalar field $\phi$ than linear scaling.}. In that case, $\alpha_\mathrm{M}$ creates anisotropic stress.
    \item \emph{kineticity} $\alpha_\mathrm{K}$\\
    This is the ``kinetic energy of scalar perturbations arising directly from the action'' \citep{bellini_maximal_2014}. This parameter is largely unconstrained by the LSS \citep{spurio_mancini_kids_2019}. 
    \item \emph{braiding} $\alpha_\mathrm{B}$\\
    This term measures the mixing between the canonical kinetic terms of the scalar field and the metric. It is particularly interesting for the scope of this {paper} because it modifies the growth of perturbations and the shape of the power spectrum, by causing ``clustering'' of dark energy\,\citep{bellini_maximal_2014}.
    \item \emph{tensor speed excess} $\alpha_\mathrm{T}$\\
    This parameter quantifies the deviation of the speed of gravitational waves $c_\mathrm{grav}$ to the speed of light $c_\mathrm{light}$. Using multi-messenger astronomy, \citet{abbott_gravitational_2017} have shown that
    \begin{align}
        \label{eq:theo_alphaM_bound}
        \left|1-c_\mathrm{grav}/c_\mathrm{light}\right| < 3 \times 10^{-15}
    \end{align}
    {on scales LIGO\,\citep{ligo_2015_advanced} is sensitive to. However, the tensor speed excess could be in principle scale dependent due to various screening mechanisms and the implications of this measurement as still debated\,\citep{rham_2018_gravitational, noller_2020_cosmological}. The Laser Interferometer Space Antenna\,(LISA, \citealt{Auclair_2023_LISA}) will likely be sensitive enough to give conclusive measurements of $\alpha_\text{T}$ on cosmic scales. Nevertheless, we employ the assumption here that $\alpha_\mathrm{T}=0$ for linear perturbations. This rules out a substantial amount of modified gravity theories acting on dark energy. Still, popular theories like quintessence\,\citep{ratra_1988_quintessence} and $f(R)$\,\citep{carroll_2004_cosmic} remain viable\,\citep{ezquiaga_2017_dark}}.
\end{itemize}
Many previously investigated modified gravity theories can be recovered if these functions take on a specific form (see \textcite{bellini_constraints_2016} for a non-exhaustive list). Most notably, for $\alpha_\mathrm{M}=\alpha_\mathrm{K}=\alpha_\mathrm{B}=\alpha_\mathrm{T}=0$ the Lagrangian reduces to the one from GR and the standard model of cosmology is retrieved. 
\\\\
In practice, we link the evolution of $\alpha_i(t)$ to the dark energy density $\Omega_\Lambda(t)$ via linear scaling coefficients $\hat{\alpha}_i$ according to
\begin{align}
    \alpha_i(t) = \hat{\alpha}_i\Omega_\Lambda(t)\;.
\end{align}
Well aware that this parameterisation can break down at redshifts $z>10$ \citep{linder_2017_challenges}, a strong motivation for Horndeski theory is whether the late time accelerated expansion of the Universe can (partly) be explained with a non-GR gravity on large cosmic scales. The energy density is driven by the same functions as the $\alpha_i$, so we could naively expect $\hat{\alpha}_i\sim\mathcal{O}(1)$ \citep{bellini_constraints_2016}. Thus, since the kineticity affects LSS evolution only marginally at best, we adapt henceforth $\hat{\alpha}_\mathrm{K}=1$. {Considering the constraints on $\alpha_\mathrm{T}$}, the Horndeski theory cosmological parameters inferred in this work will be the braiding and Planck mass run rate proportionality constants, $\hat{\alpha}_\mathrm{B}$ and $\hat{\alpha}_\mathrm{M}$, respectively. Note that we do not vary the dark energy equation of state here explicitly since our focus is on the modifications to GR on the perturbation level and not at the background.
\\\\
Naturally, changing the field equations that determine the metric also affect geodesics, influencing the weak gravitational lensing formalism in particular. We choose conformal Newtonian gauge for the coupling between geodesics and matter fields which is a common choice for scalar perturbation theories \citep{ma_cosmological_1995}:
\begin{align}
    \label{eq:theo_line_element}
    \mathrm{d}s^2=a^2(\tau)\left[-c^2\left(1+\frac{2\Psi}{c^2}\right)\mathrm{d}\tau^2+\left(1-\frac{2\Phi}{c^2}\right)\mathrm{d}\boldsymbol{x}^2\right] \; .
\end{align}
$c$ is the speed of light and $\Psi$ and $\Phi$ are the Newtonian perturbative potentials, also known as Bardeen potentials \citep{bardeen_gauge-invariant_1980}, dependent on the position $\boldsymbol{x}$ and scale factor $a$. Following \citet{planck_collaboration_xiv_planck_2015}, we introduce two phenomenological functions in Fourier space that characterise linear structure formation in modified gravity: The ratio of the Bardeen potentials 
\begin{align}
    \label{eq:theo_bardeen_ratio_realspace}
    \eta(\boldsymbol{k},a) := \frac{\tilde{\Phi}(\boldsymbol{k},a)}{\tilde{\Psi}(\boldsymbol{k},a)}
\end{align}
and the change in Poisson equation $\mu(\boldsymbol{k},z)$ defined by 
\begin{align}
    \label{eq:theo_poisson_mod}
    -k^2\tilde{\Psi}(\boldsymbol{k},a) = 4\pi G a^2\tilde{\rho}_\mathrm{m}(\boldsymbol{k},a)\mu(\boldsymbol{k},a)
\end{align}

\begin{figure}
    \centering
    \includegraphics[width=1\linewidth]{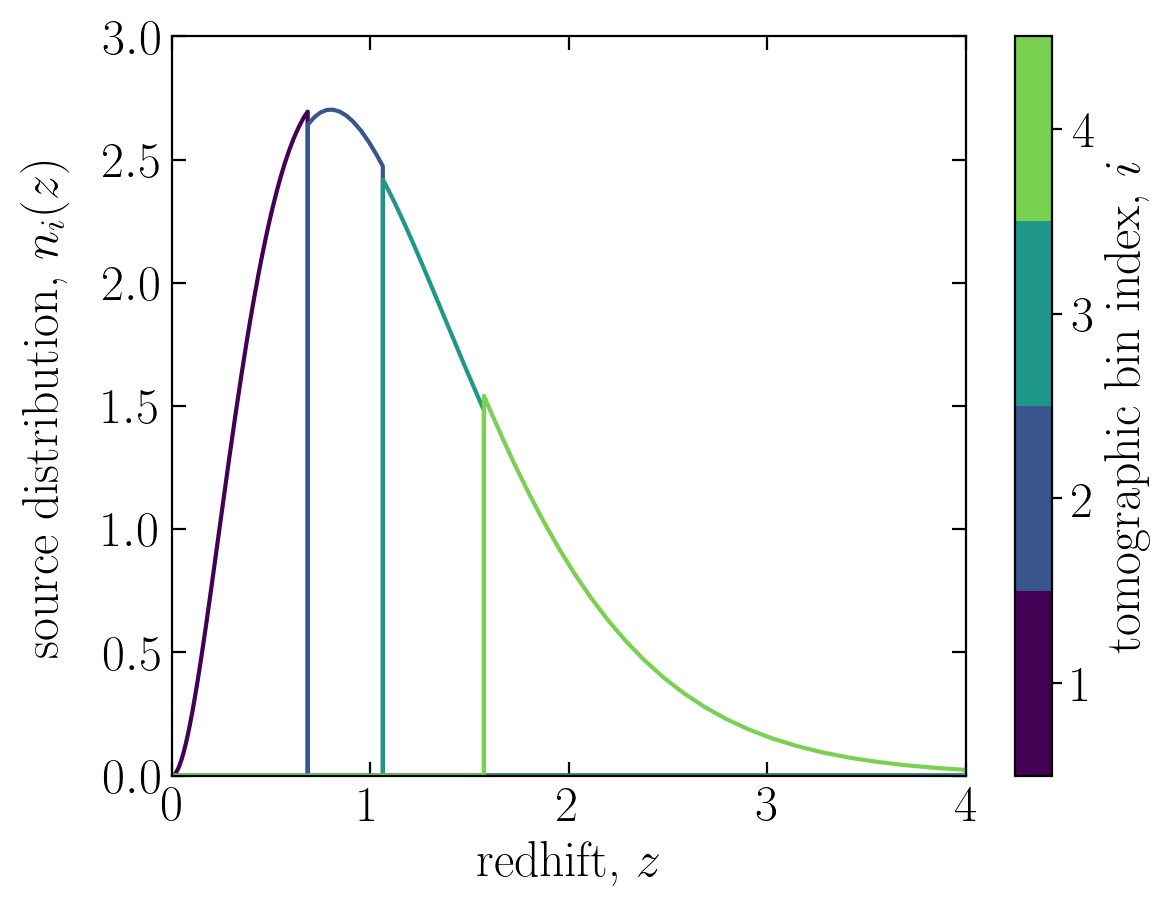}
    \caption{Tomographic source redshift distribution of the mock FRB catalogue. {The discontinuities between bins originate from normalising each source distribution individually.}}
    \label{fig:FRB_redshift_dist}
\vspace{.25cm}\end{figure}

with the Fourier transformed matter density $\tilde{\rho}_\mathrm{m}(\boldsymbol{k},a)$ and wave-number $k:=|\boldsymbol{k}|$. Note that for ${\mu},{\eta}\rightarrow 1$, GR is retrieved. {We know that GR is an excellent description of gravity for very small scales at the very least\,\citep{wojtak_gravitational_2011}, hence we force modified gravity to satisfy GR on these scales by the means of screening. There are a variety of proposed screening mechanisms (see \citealt{joyce_beyond_2015} for an excellent review thereof). The ones most commonly discussed in literature are Chameleon\,\citep{khoury_2004_chameleon, khoury_chameleon_2004-1} and Vainshtein screening\,\citep{vainshtein_problem_1972, babichev_introduction_2013}. In Chameleon-like screening, the mass of the scalar field depends on the local matter density, making the effects of modified gravity for high-density environments like Earth short-ranged, hence the name. On the other hand, Vainshtein screening suppresses scalar field effects in the presence of massive sources where the $\phi$-kinetic terms becomes large. This happens on scales smaller than a characteristic distance named the Vainshtein radius, which is dependent on the properties of the gravity source. In this work, we employ a phenomenological model described in \citet{reischke_investigating_2019,spurio_mancini_kids_2019}, where $\eta(k, a)$ and $\mu(k, a)$ decay exponentially to unity for scales smaller than the screening scale $\lambda_\mathrm{s}\sim 1/k_\mathrm{s}$. In this work, we refrain from testing the effects of implementing a variety of screening mechanisms. However, preliminary results suggest that the approach taken here does not influence the power spectrum up to a few per cent\,(Grasso et al., in prep.).}

\begin{figure}
    \centering
    \includegraphics[width=1\linewidth]{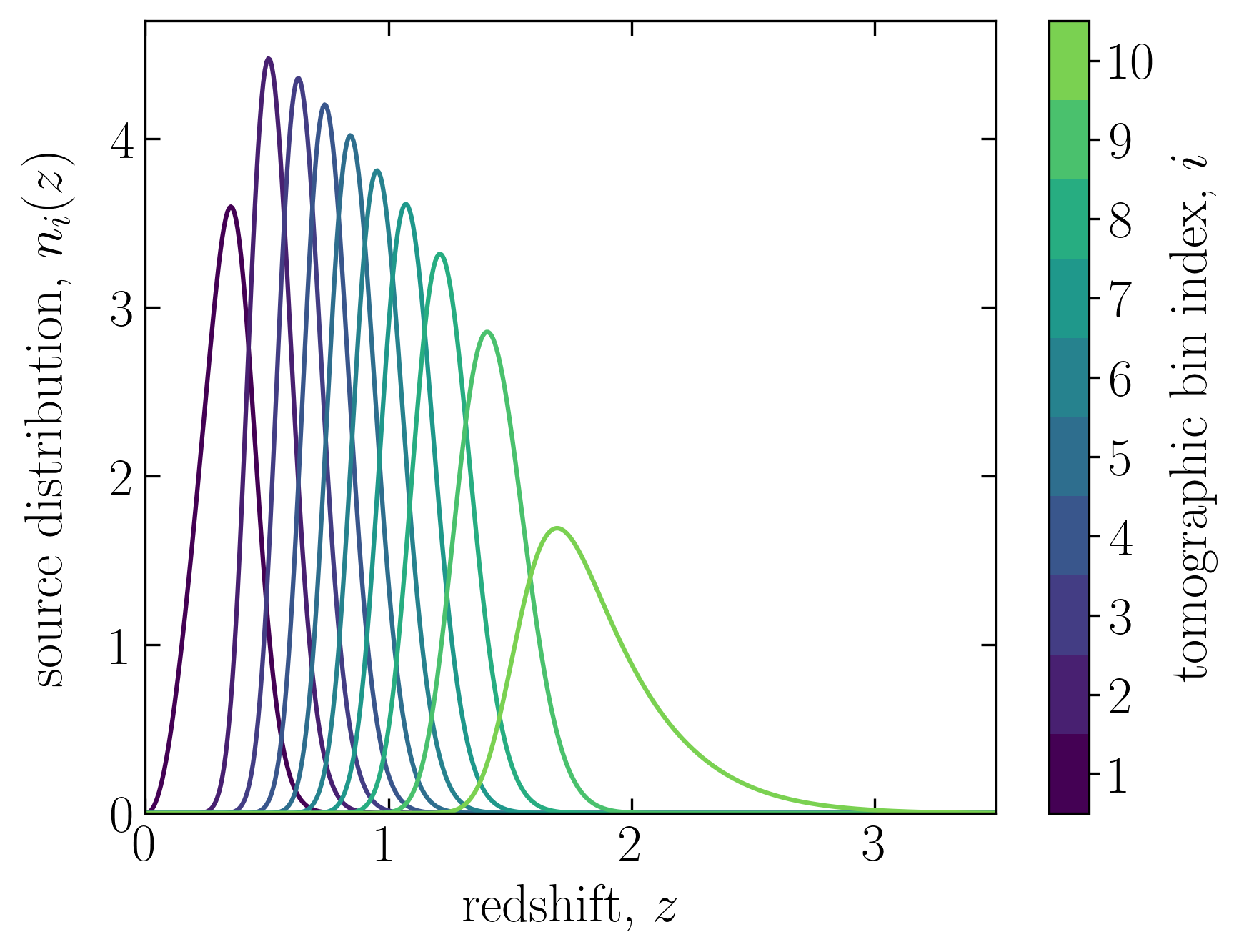}
    \caption{Tomographic source redshift distribution for a stage-4 cosmic shear survey. Adapted to match the \textsc{Euclid-CS} forecast \citep{blanchard_euclid_2020}.}
    \label{fig:Euclid_redshift_dist}
\vspace{.25cm}\end{figure}

{One would expect the screening scale to affect extremely non-linear scales and/or high power spectrum amplitudes with the Chameleon or Vainshtein screening. However, we restrict the screening scale to be below intermediate scales, $k_\mathrm{s}<2\,\mathrm{Mpc}^{-1}$ with a fiducial value at $k_\mathrm{s} = 0.1\,\mathrm{Mpc}^{-1}$. The reason for only considering (semi-)linear scales is of practical nature: We use \texttt{HMcode} to model the non-linear power spectrum (see also Section\;\ref{ch:theory_3dpower_spec}). Notably, \texttt{HMcode} was developed and calibrated for a $\Lambda$CDM cosmology and has no inherent knowledge and functionality of modified gravity. While there are efforts to model the non-linear matter power spectrum in scalar-tensor perturbation theory\,\citep{cantaneo_2019_on, bose_2023_fast}, there is little similar quantitative research for the baryon power spectrum. To circumvent this issue we remove modified gravity for large $k$, avoiding the necessity of modelling the effect of modified gravity on baryons. Therefore, including modified gravity on non-linear scales would only increase information gain therein, leaving the results of this work valid as a lower bound.}

\subsection{Angular power spectrum}

We assume the spherical harmonic coefficients $\hat{a}_{\ell m}$ of the fields of interest in this work to follow a zero-mean Gaussian distribution. The measured coefficient consists of a signal ($a_{\ell m}$) and a noise term ($n^{(a)}_{\ell m}$) such that
\begin{align}
    \hat{a}_{\ell m} = a_{\ell m} + n^{(a)}_{\ell m}\; .
\end{align}
Consequently, the angular power spectrum of two fields with coefficients $\hat{a}_{\ell m}$, $\hat{b}_{\ell m}$ can be split into signal and noise as well:
\begin{align}
    \label{eq:theo_measured_Cell}
    \hat{C}^{AB}_\ell = C_\ell^{AB} + N_\ell^{AB}\delta^\mathrm{K}_{AB}\;,
\end{align}
where we defined the angular power spectrum as the non-geometrical component of the correlator such that: 
\begin{align}
    \hat{C}^{AB}_\ell\delta_{\ell \ell'}\delta_{m m'} = \langle \hat{a}_{\ell m}\hat{b}^*_{\ell' m'} \rangle\;,
\end{align}
for statistically isotropic and homogeneous fields. The Kronecker-Delta, $\delta^\mathrm{K}_{AB}$, appears since we assume the noise contributions of the FRB and WL measurement to be uncorrelated. The noise term in Eq.\;\ref{eq:theo_measured_Cell} is specific to the type of measurement. However, the signal can be more generally expressed as the projection of the three-dimensional power spectrum $P^{AB}(k,a)$ with the appropriate kernel(s) $W_{A/B}(\chi)$ via the Limber approximation,\citep{limber_analysis_1954} which is sub-percent accurate for $\ell>10$ \citep{kilbinger_precision_2017}. Including tomographic redshift bins $i,\,j$, it can be written as 
\begin{align}
    \label{eq:theo_cell_limber}
    \left(C^{AB}_\ell\right)_{ij} = \int_0^{\chi_\mathrm{H}}\!\!\mathrm{d}\chi\, \frac{W^{(i)}_{A}(\chi)  W^{(j)}_{B}(\chi)}{\chi^2} \, P_\mathrm{AB}\!\left(\frac{\ell+\frac{1}{2}}{\chi}, z(\chi)\right)
\end{align}
with $\chi_\mathrm{H} := \chi(z\rightarrow\infty)$ as the comoving horizon size\footnote{Not to be confused with the Hubble distance $\chi_{\mathrm{H}_0}=c/\mathrm{H}_0$}.

\begin{figure}
    \centering
    \includegraphics[width=1\linewidth]{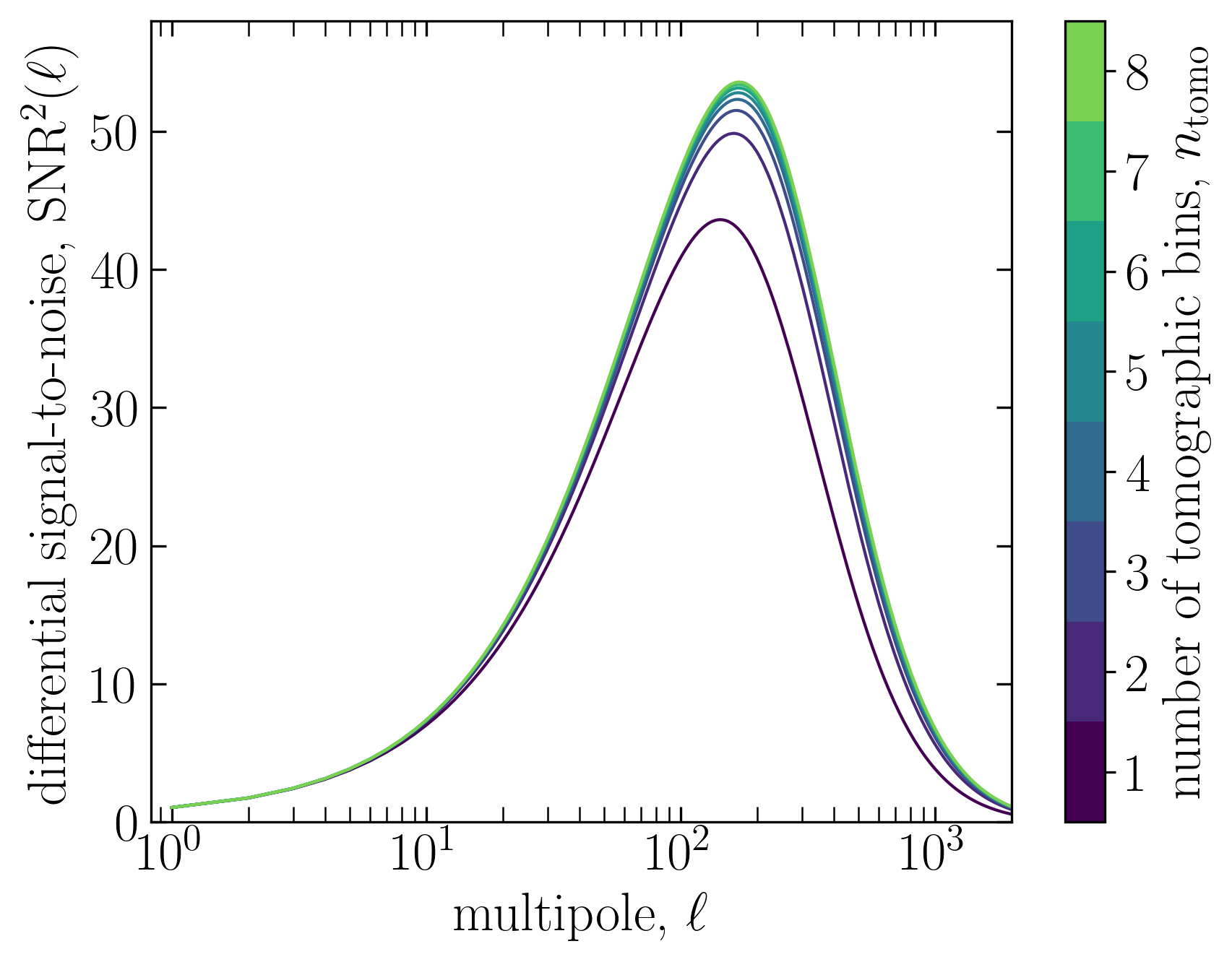}
    \caption{Squared differential SNR dependent on angular scale of the FRB angular power spectrum for different tomographic bin numbers shown as a colour bar.}
    \label{fig:data_SNR_of_FRB_tomo_z}
\vspace{.25cm}\end{figure}

\subsubsection{Weak lensing statistics}

Shear, commonly denoted with $\gamma$, is defined as the trace-free part of the matrix that maps the local light distribution onto the observer image in the presence of a lensing potential $\psi$. Using a tomographic source distribution $n_{i,\gamma}(z)$, the corresponding kernel (or window function) in GR is given by
\begin{align}
    W^{(i)}_{\gamma,\mathrm{GR}}(\chi) = \frac{3H_0^2}{2c^2} \Omega_{\mathrm{m}0} \frac{\chi}{a(\chi)}\int_{\chi}^{\chi_\mathrm{H}}\mathrm{d}\chi' \,\frac{\chi'-\chi}{\chi'}n_{i,\gamma}(\chi')\;,
\end{align}
where $H_0$ is the Hubble constant, $\Omega_{\mathrm{m}0}$ the matter density parameter today and $n_{i,\gamma}(\chi')\mathrm{d}\chi' = n_{i,\gamma}(z)\mathrm{d}z$ \citep{kilbinger_cosmology_2015}\footnote{The transformation $n_{i}(\chi)\mathrm{d}\chi = n_{i}(z)\mathrm{d}z$ holds for the DM source distribution as well.}. However, gravity that does not conform with GR naturally affects the distortion of light through WL in a different way. We can absorb non-GR terms originating from matter overdensities into the shear window function, such that
\begin{align}
    \label{eq:theo_wl_window_mod}
    W^{(i)}_{\gamma,\mathrm{mod}}(k,\chi) = \frac{1}{2}\mu(k,\chi)\left(1+\frac{1}{\eta(k,\chi)}\right) W^{(i)}_{\gamma,\mathrm{GR}}(\chi)\;.
\end{align}
See \citet{spurio_mancini_kids_2019} for a more detailed derivation, but be aware of the different naming conventions of the Bardeen potentials. {The quasi-static approximation (QSA) is often used in the perturbative Einstein equations, where time derivatives are neglected, allowing for an explicit relationship between the $\alpha_i$ and $\mu(k,\chi)$ as well as $\eta(k,\chi)$\citep{sawicki_2015_limits}. However, the \texttt{hi\_class} modification we use (see also Ch. \ref{ch:theory_3dpower_spec}) does not employ the QSA. Instead, the density and potential are sourced directly from \texttt{hi\_class} that considers the full statistical properties of the Weyl potential\,\citep{spurio_mancini_testing_2018}.}
Inserting Eq.\;\ref{eq:theo_wl_window_mod} with $k=\frac{\ell+1}{2}$ into Eq.\;\ref{eq:theo_cell_limber} with the matter power spectrum $P_\mathrm{mm}$ yields the auto-correlated cosmic shear angular power spectrum matrix in modified gravity.

The noise component of the observed shear spectra, Eq.\;\ref{eq:theo_measured_Cell}, called shape noise is given as
\begin{align}
	\label{eq:theo_shear_noise_contribution}
    \left(N^{\gamma}\right)_{ij} = \left(\frac{\sigma_{\epsilon,i}/\sqrt{2}}{\sqrt{\bar{n}_{i,\gamma}}}\right)^2\delta^\mathrm{K}_{ij}\; .
\end{align}
$\sigma_{\epsilon,i}$ is the total intrinsic ellipticity dispersion of sources in tomographic bin $i$. Typical values are around 0.3
\citep[e.g.][]{laureijs_euclid_2011,blanchard_euclid_2020}. $\bar{n}_{i,\gamma}$ is the number of sources per square radiant in tomographic bin $i$. The shape noise from different redshift bins is uncorrelated as they stem from different sources, and, therefore, the noise contribution is assumed to be diagonal, expressed via $\delta^\mathrm{K}_{ij}$.

\begin{table}
	\centering
 \renewcommand{\arraystretch}{1.15}
	\begin{tabular}{lcc}
		\hline
		\multicolumn{1}{c}{parameter} & description & value  \\ \hline
		$n_{\epsilon,\mathrm{tomo}}$ & tomographic source bins & 10 \\ 
  $n_{\mathrm{tomo}}$ & tomographic DM bins & 4 \\ 
        $\sigma_{\epsilon,i}$ & ellipticity dispersion & 0.3 \\
        $\sigma_{\mathrm{host}0}$ $[\mathrm{pc}\,\mathrm{cm}^{-3}]$ & DM dispersion & 50 \\
        $\bar{n}_{\epsilon,i}\,[\mathrm{arcmin}^{-2}]$ & source density & 3 \\
        $\bar{n}_{\mathrm{FRB},i}\,[\mathrm{arcmin}^{-2}]$ & FRB density & 0.086 \\
        $f^\gamma$ & source sky footprint & 0.3 \\
        $f^\mathcal{D}$ & FRB sky footprint & 0.7 \\  
        $\ell_\mathrm{max}$ & maximum angular scale & 5000 \\
    \end{tabular}
    \caption{Survey specifications for the mock FRB and stage-4 cosmic shear survey used in the forecast presented here.}
    \label{tab:Euclid_specs}
\end{table}

\subsubsection{Dispersion measure statistics}

\begin{table*}
\centering
 \renewcommand{\arraystretch}{1.4}
    \begin{tabular}{clcccc}
    \hline
        \phantom{aa}& \multicolumn{1}{c}{Parameter} & description & fiducial value & prior range & source\\
         \hline
        \parbox[t]{2mm}{\multirow{10}{*}{\rotatebox[origin=c]{90}{investigated parameters}}} & $\hat{\alpha}_\mathrm{B}$ & braiding function proportionality & $0.05$ & [$0$, $2.5$] & \cite{spurio_mancini_testing_2018}\\
        & $\hat{\alpha}_\mathrm{M}$ & Planck mass run rate proportionality & $0.05$ & [$0$, $3$] & \cite{spurio_mancini_testing_2018}\\
        & $\log_{10}(\frac{k_\mathrm{s}}{\mathrm{Mpc}^{-1}})$ & Horndeski theory screening scale & -1 & [$-2$, $0.3$] & \cite{spurio_mancini_testing_2018} \\
        & $h$ & dimensionless Hubble parameter & 0.674 & [$0.38$, $1.00$] & \cite{planck_collaboration_planck_2020}\\
        & $\Omega_\mathrm{b}$ & baryon density parameter & $0.04931$ & [$0.015$, $0.1$] & \cite{planck_collaboration_planck_2020}\\
        & $\Omega_\mathrm{cdm}$ & cold dark matter density parameter & $0.2642$ & [$0.18$, $0.34$] & \cite{planck_collaboration_planck_2020}\\
        & $n_\mathrm{s}$ & primordial scalar index & $0.965$ & [$0.7$, $1.25$] & \cite{planck_collaboration_planck_2020}\\
        & $\sum m_\nu/\mathrm{eV}$ & total mass of all neutrino species & $0.06$ & [$0.003$, $1.5$] & \cite{planck_collaboration_planck_2020}\\
        & $\sigma_8$ & linear power spectrum RMS fluctuations & $0.811$ & [$0.7$, $0.92$] & \cite{planck_collaboration_planck_2020}\\
        & $\log_{10}(\frac{T_\mathrm{AGN}}{\mathrm{K}})$ & baryonic feedback strength & $7.8$ & [$7.0$, $8.6$] & \cite{mead_hydrodynamical_2020} \\ \hline
        \parbox[t]{2mm}{\multirow{10}{*}{\rotatebox[origin=c]{90}{fixed values}}}& $N_\nu$ & number of \textsb{equal mass} neutrino species & 3 & & \cite{planck_collaboration_planck_2020}\\
        & $M_\mathrm{Pl}$ & initial Planck mass & $M_\mathrm{Pl}^\mathrm{obs}$ & & - \\
        & $\hat{\alpha}_\mathrm{K}$ & kineticity function proportionality & $1$ & & - \\
        & $\hat{\alpha}_\mathrm{T}$ & tensor excess speed function proportionality & $0$ & &\citet{abbott_gravitational_2017} \\
        & $N_\mathrm{eff}$ & effective relativistic degrees of freedom & 3.046 & &\cite{planck_collaboration_planck_2020}\\
        & $T_0$ & Black body CMB temperature today & $2.7255\,\mathrm{K}$ & &\citet{fixsen_temperature_2009}\\
        & $z_\mathrm{reio}$ & reionisation redshift & 7.7 & &\cite{planck_collaboration_planck_2020}\\
        & $Y_\mathrm{He}$ & primordial helium fraction & 0.246 & &\cite{planck_collaboration_planck_2020}\\
        & $w(a)$ & dark energy equation of state & -1 & &\cite{planck_collaboration_planck_2020}\\
        & $\Omega_\mathrm{K}$ & curvature density parameter & 0 & &\cite{planck_collaboration_planck_2020}\\ 
    \end{tabular}
        \caption{Table with the adapted fiducial cosmology and the prior ranges for the different parameters used in this analysis.}
    \label{tab:data_fid_cosmology}
    \vspace{.25cm}
\end{table*}

An FRB is a broadband pulse that experiences dispersion when travelling through the intergalactic medium. The time delay of frequency $\nu$ accumulated by propagating from source at $l=L$ to observer at $l=0$ can be measured as $\Delta t \propto \nu^2$, where (up to some physical constants) the proportionality constant is the integrated electron column density along the line of sight $\mathrm{d}l$, also known as the dispersion measure
\begin{align}
    \mathrm{DM} = \int_0^L \mathrm{d}l\, n_\mathrm{e}(l) \;.
\end{align}
The dispersion caused by the LSS is blended in with dispersion accumulated from the Milky Way (MW) and host galaxy halo where the FRB progenitor resides. Hence, the total DM dependent on angular sky position $\hat{n}$ and redshift $z$ can be written as the sum of its components \citep{Mo_2022_dispersion}:
\begin{align}
    \mathrm{DM}_\mathrm{tot}(\hat{n},z) = \mathrm{DM}_\mathrm{MW}(\hat{n}) + \mathrm{DM}_\mathrm{LSS}(\hat{n},z) + \mathrm{DM}_\mathrm{host}(z) \; .
\end{align}
The LSS contribution can be further split into an isotropic and anisotropic part:
\begin{align}
    \label{eq:theo_DM_fluc_definition}
    \mathrm{DM}_\mathrm{LSS}(\hat{n},z) = \langle \mathrm{DM}_\mathrm{LSS} \rangle(z) + \mathcal{D}(\hat{n},z) \;.
\end{align}
In this work, we investigate DM fluctuations $\mathcal{D}(\hat{n},z)$ which enables us to employ a zero-mean Gaussian field assumption. The corresponding auto-correlated angular power spectrum is then retrieved from Eq.\;\ref{eq:theo_cell_limber} with the electron power spectrum $P_\mathrm{ee}$ and the DM kernel
\begin{align}
    \label{eq:theo_DM_kernel}
    W^{(i)}_\mathcal{D}(\chi) = \mathcal{A}F(z(\chi))(1+z(\chi))\int_\chi^{\chi_\mathrm{H}}\mathrm{d}\chi'\, n_{i,\mathcal{D}}(\chi')
\end{align}
with
\begin{align}
    \mathcal{A} = \frac{3 H_0^2 \Omega_{\mathrm{b}0}}{8\pi G m_\mathrm{p}}\;,
\end{align}
that contains the dimensionless baryon density parameter today $\Omega_{\mathrm{b0}}$, the gravitational constant $G$, the proton mass $m_\mathrm{P}$ and the tomographic FRB source distribution $n_{i,\mathcal{D}}(\chi)$. For the fraction of free electrons in the intergalactic medium, $F(z)$, one needs to take into account the fraction of ionised electrons of hydrogen $X_{\mathrm{e},\mathrm{H}}(z)$ and helium $X_{\mathrm{e},\mathrm{He}}(z)$ with their respective primordial mass fractions $Y_\mathrm{H}$, $Y_\mathrm{He}$. Using the total fraction of baryons in the intergalactic medium $f_\mathrm{IGM}(z)$, this can be written as
\begin{align}
    \label{eq:theo_electron_IGM_fraction_full}
    F(z) = f_\mathrm{IGM}(z)\left[Y_\mathrm{H}X_{e,\mathrm{H}}(z)+\frac{1}{2}Y_\mathrm{He}X_{\mathrm{e},\mathrm{He}}(z)\right] \;.
\end{align}

However, for the considered redshift range in this work ($z\leq4$), the intergalactic medium can be assumed to be fully ionised \citep{aghanim_2001_reionisation}. Additionally, the majority of the DM signal comes from $0.5<z<2$ (see also the FRB mock source distribution in Section \ref{ch:frb_catalogue}) at which we assume a constant $f_\mathrm{IGM}\approx0.9$. The exact distribution of baryons is still an object of research \citep{shull_baryon_2012} and a more complex model can be implemented in future work. Combining this assumption with the vanishing fraction of heavy elements ($Y_\mathrm{H}\approx 1- Y_\mathrm{He}$), Eq.\;\ref{eq:theo_electron_IGM_fraction_full} reduces to
\begin{align}
    \label{eq:theo_electron_IGM_fraction_approx}
    F(z) \approx F_\mathrm{e} = 0.9 \left[1-\frac{1}{2}Y_\mathrm{He}\right]\;.
\end{align}
A proper model of the MW and host galaxy impact in terms of noise is equally important. Extensive work has been done on the contribution of the MW \citep{cordes_new_2002,yao_new_2017,yamasaki_galactic_2020}
and we assume here that it can be accurately modelled and subtracted. {In reality this might not be the case and improper $\text{DM}_\text{MW}$ modelling will lead to large-scale contributions of the DM-DM auto-correlation. However, since shear and $\text{DM}_\text{MW}$ are uncorrelated, we expect the gain in constraining power that is driven by the cross-correlation largely unaffected. Furthermore, selecting FRBs outside of the MW disk would reduce the possible residual non-LSS DM contributions.} Meanwhile, the host contribution acts as a Poisson-like flat noise contribution that can be characterised by the host DM dispersion
\begin{align}
    \label{eq:theo_DM_host_variance}
\sigma_\mathrm{host}(z) = \frac{\sigma_\mathrm{host0}}{1+z} \approx \frac{50\,\mathrm{pc}\,\mathrm{cm}^{-3}}{1+z} . 
\end{align}
The factor $1+z$ originates from the transformation between the FRB rest- and observer frame \citep{Zhang_2020_dispersion}. {The host dispersion is still a large source of uncertainty\,\citep{Tang_2023_inferring}. Simulations suggest that the $\sigma_\text{host0}$ is rather on order of $\sim$100\,pc\,cm$^{-3}$\,\citep{kovacs_2024_dispersion, zhang_2024_BINGO}. Still, here we fix the contribution to the observationally motivated value from \citet{arcus_2021_fast}. Even though the assumed host dispersion measure might be optimistic, it is highly degenerate with the total number of FRBs\,\citep[Fig. 4 therein]{reischke_calibrating_2023}. Recent estimates by \citet{connor_2023_stellar} suggest that the supposed figure of 10000 FRBs until the end of the decade might be on the low end. Thus, we do not expect our choice of parameters to affect the conclusion of this paper. A more rigorous approach would be to additionally marginalise over the host DM dispersion with an analytical model, like we propose in \citet{reischke_2024_analytical}. However, this is outside the scope of this paper, hence we choose a simplistic host dispersion model. As such, we can write down the noise contribution as}

\begin{align}
    (N^{\mathrm{DM}})_{ij} = \left(\frac{\sigma_\mathrm{host}(\bar{z}_i)}{\sqrt{\bar{n}_{i,\mathrm{FRB}}}}\right)^2\delta_{ij}\;,
\end{align}
where $n_{i,\mathrm{FRB}}$ is the effective number of FRBs per solid angle and
\begin{align}
	\bar{z}_i = \frac{\int_{z_i}^{z_{i+1}}\mathrm{d}z\, n_i(z)  z}{\int_{z_i}^{z_{i+1}}\mathrm{d}z\, n_i(z)} \;.
\end{align}
is the weighted mean redshift in bin $i$. With the formalisms above, the cross-correlated angular power spectrum can be easily obtained by inserting the matter-electron power spectrum
\begin{align}
    \label{eq:matter_electron_pk}
    P_\mathrm{em}(k,z) = \sqrt{P_\mathrm{ee}(k,z)  P_\mathrm{mm}(k,z)}
\end{align}
in addition to WL (Eq.\;\ref{eq:theo_wl_window_mod}) and DM (Eq.\;\ref{eq:theo_DM_kernel}) kernel into Eq.\;\ref{eq:theo_cell_limber}.

\section{Methodology}
\label{ch:method}

\subsection{MCMC likelihood}
We forecast the uncertainties on the fiducial set of cosmological parameters $\{\hat{\theta}\}_i$ with a Markov-Chain-Monte-Carlo (MCMC) algorithm. We denote the fiducial spherical harmonic vector as $\hat{\boldsymbol{a}}_{\ell m}$, with the corresponding power spectrum as $\hat{\boldsymbol{C}}_\ell$. Assuming a Gaussian likelihood $L$ with covariance $\hat{\boldsymbol{C}}_\ell$, we can express the expected log-likelihood of measuring $\boldsymbol{a}_{\ell m}$, given that $\hat{\boldsymbol{a}}_{\ell m}$ are the true harmonics, as
\begin{align}
\label{eq:data_log_likelihood}
\langle \mathcal{L} \rangle = & \; \langle-\ln(L(\boldsymbol{a}_{\ell m}|\boldsymbol{\mu}=0, \hat{\boldsymbol{C}}_\ell))\rangle \\
= & \; f_{\mathrm{sky}}\sum_\ell \frac{2\ell+1}{2} \mathrm{Tr}\!\left(\ln(\hat{\boldsymbol{C}}_\ell)+\hat{\boldsymbol{C}}_\ell^{-1}\boldsymbol{C}_\ell\right) \; ,
\end{align}
where Tr is the trace-operator, $\hat{\boldsymbol{C}}_\ell^{-1}$ is the inverse of the covariance and the factor $(2\ell+1)$ originates from the independence of the $m$-modes which take all integers in the interval $m\in[-\ell,\ell]$ \citep{tegmark_karhunen-loeve_1997}. $f_\mathrm{sky}$ is a multiplicative factor that effectively decreases the amount of measured $m$-modes due to partial sky coverage of the used survey \citep{leistedt_estimating_2013,alonso_unified_2019,nicola_cosmic_2021}.
For a realistic cross-correlation measurement, one has to take into account the effects of differing sky fractions. As an applicable example, consider a measurement of cosmic shear $\gamma$ and DM fluctuations $\mathcal{D}$ with sky fraction $f^{\gamma}, \,f^{\mathcal{D}}$ respectively, whereas $f^{\gamma}$ is smaller and fully contained in $f^{\mathcal{D}}$. Furthermore, the full-sky cross-correlation likelihood is calculated with the auto- and cross-correlated angular power spectra matrices 
\begin{align}
	\label{eq:theo_cross_angular_power_spec}
    \boldsymbol{C}_\ell = 
    \begin{pmatrix}
        \boldsymbol{C}_\ell^{\mathcal{DD}} & \boldsymbol{C}_\ell^{\mathcal{D}\gamma} \\[0.5em]
        \boldsymbol{C}_\ell^{\gamma\mathcal{D}} & \boldsymbol{C}_\ell^{\gamma\gamma} 
    \end{pmatrix}
\end{align}
that contain the tomographic angular power spectra defined above. Similarly to our previous analysis \citep{reischke_calibrating_2023}, the cross-correlated likelihood must be calculated on the smaller footprint $f^{\gamma}$ and the contribution from the remaining sky fraction $(f^{\mathcal{D}}-f^{\gamma})$ is manually added:
\begin{align}
    \langle \mathcal{L} \rangle^{\mathcal{D}\gamma} = f^{\gamma}\langle \mathcal{L} \rangle^{\mathcal{D}\gamma}_\mathrm{full\,sky} + (f^{\mathcal{D}}-f^{\gamma})\langle \mathcal{L} \rangle^{\mathcal{D}}_\mathrm{full\,sky} \; .
\end{align}
Note that $\langle \mathcal{L} \rangle_\mathrm{full\,sky}$ is calculated from Eq.\;\ref{eq:data_log_likelihood} with $f_\mathrm{sky}=1$.

\begin{table*}
 \renewcommand{\arraystretch}{1.75}
    \centering
    \begin{tabular}{lcccc}
    \hline
        \multicolumn{1}{c}{Parameter} & FRB & \textsc{Euclid-CS} & \textsc{Euclid-CS}$\times$FRB & reduction factor for $\hat{\alpha}^{\mathrm{fid}}_\mathrm{B/M}=0.05$ ($\hat{\alpha}^{\mathrm{fid}}_\mathrm{B/M}=1$) \\ \hline
        $\Omega_\mathrm{b}$ & $0.0455^{+0.0056}_{-0.011} $ & $0.046^{+0.012}_{-0.010}   $ & $0.0484^{+0.0028}_{-0.0025}$ & {4.2 (4.7)} \\ 
        $\Omega_\mathrm{cdm}$ & $0.245\pm 0.022            $ & $0.261^{+0.010}_{-0.012}   $ & $0.2625\pm 0.0032          $ & {3.5 (3.2)} \\ 
        $h$ & $0.78^{+0.14}_{-0.12}      $ & $0.838^{+0.15}_{-0.060}    $ & $0.686^{+0.030}_{-0.043}   $ & 2.9 (2.9) \\ 
        $n_\mathrm{s}$ & $0.97\pm 0.13              $ & $0.864\pm 0.072            $ & $0.956^{+0.037}_{-0.030}   $ & 2.2 (2.2) \\ 
        $\sum m_\nu/\mathrm{eV}$ & $< 0.817                   $ & $< 0.647                   $ & $< 0.132                   $ & 5.0 (4.1) \\ 
        $\log_{10}\!\left(\frac{T_\mathrm{AGN}}{\mathrm{K}}\right)$ & $7.821^{+0.058}_{-0.065}   $ & $7.75^{+0.22}_{-0.18}      $ & $7.7995^{+0.0090}_{-0.0070}$ & 24.8 (16.8) \\ 
        $\sigma_8$ & $< 0.812                   $ & $0.8071^{+0.0050}_{-0.0059}$ & $0.8120^{+0.0044}_{-0.0039}$ & 1.3 (1.4) \\ 
        $\hat{\alpha}_\mathrm{B}$ & $< 1.53                    $ & $0.76^{+0.23}_{-0.73}      $ & $< 0.641                   $ & 1.5 (1.7) \\ 
        $\hat{\alpha}_\mathrm{M}$ & ---                          & $1.37^{+0.54}_{-1.3}       $ & $< 1.35                    $ & 1.3 (1.5) \\ 
        $\log_{10}\!\left(\frac{k_\mathrm{s}}{\mathrm{Mpc}^{-1}}\right)$ & ---                          & $< -1.52                   $ & $< -1.60                   $ & 1.2 (1.7) \\ 
    \end{tabular}
     \caption{Posterior mean with 68\% confidence intervals for the auto- and cross-correlations of DM and shear fluctuations with the fiducial values presented in Tab \ref{tab:data_fid_cosmology}. The reduction factor relates to the improvement from \textsc{Euclid-CS}$\times$FRB as opposed to \textsc{Euclid-CS} alone and is shown for the case $\hat{\alpha}^{\mathrm{fid}}_\mathrm{B/M}=0.05$ and $\hat{\alpha}^{\mathrm{fid}}_\mathrm{B/M}=1$ in brackets in the last column.}
    \label{tab:res_alpha_0.05}
\end{table*}

\subsection{Mock catalogue specification}

In the following, we describe the FRB and WL galaxy source distributions adapted for this work.

\subsubsection{FRB catalogue}
\label{ch:frb_catalogue}

\begin{figure}
    \centering
    \includegraphics[width=1\linewidth]{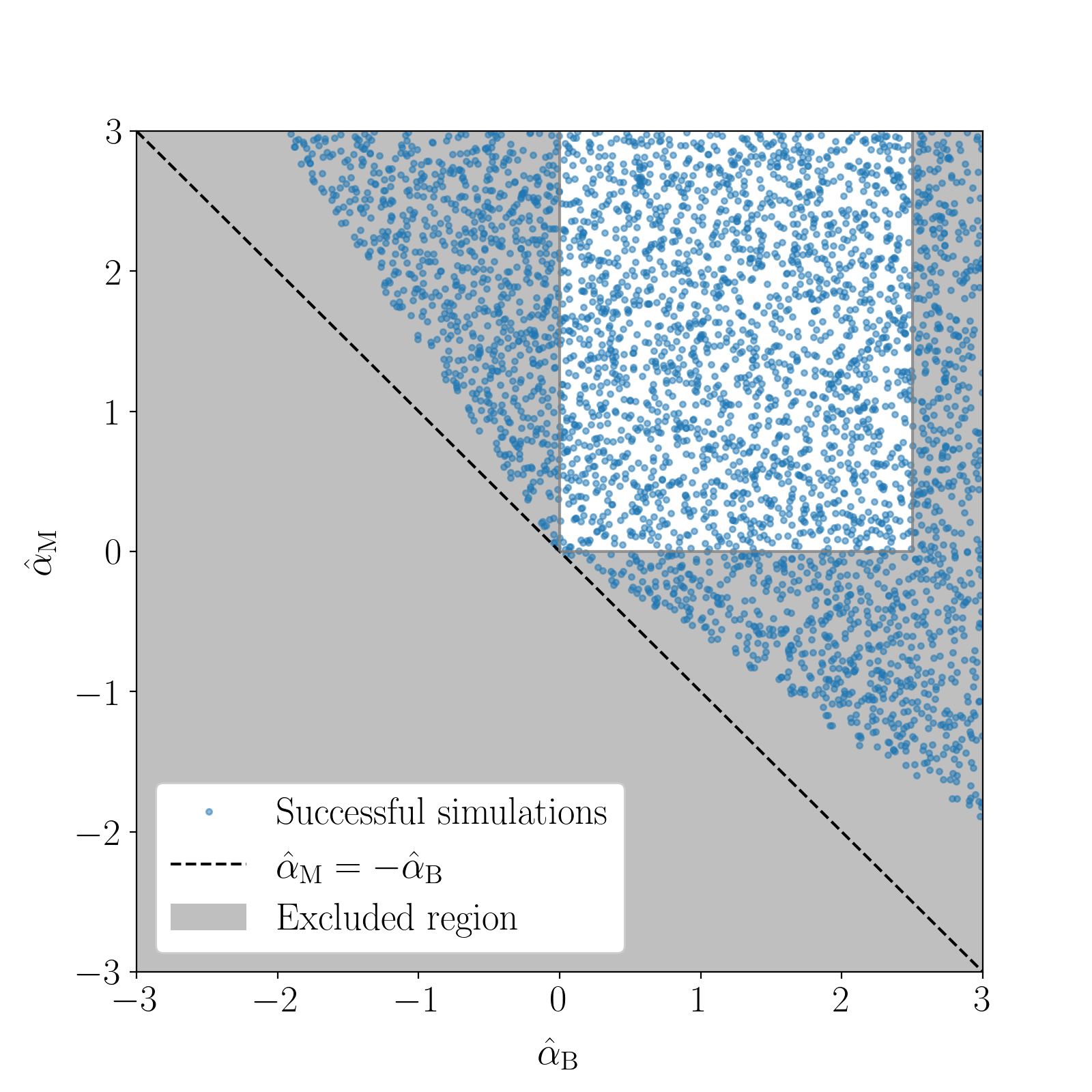}
    \caption{Projection of successful initialisations of \texttt{hi\_class} from the 11-dimensional hyperspace onto the $\aM$-$\aB$-plane. The blue dots depict the 3881 parameter sets out of 10000 candidates spaced in a Latin Hypercube that did not result in a stability condition break.}
    \label{fig:res_aBM_exploration}
\vspace{.25cm}\end{figure}
We model the FRB catalogue properties according to a survey that is feasible within the next decade based on the predictions from \citet{petroff_fast_2019}. {The actual number of host identified FRBs is rapidly growing. They are collected in the comprehensive \texttt{Blinkverse}-database\footnote{\url{https://blinkverse.zero2x.org/}}\,\citep{xu_2023_blinkverse}. As of the May 2025, there are 92 host identified FRBs.} For this analysis, we choose $10^4$ FRB samples with associated redshift, whereas a substantial fraction is located above redshift 2. Assuming the FRBs trace the galaxy distribution \citep{reischke_probing_2021}, the source distribution can be written as (assuming a flux limited sample)
\begin{align}
    n(z) \propto z^2\exp(-\alpha z) \;,
\end{align}
where $\alpha$ denotes the survey depth. Henceforth, we use $\alpha=2.5$ which fullfils the demands above. Note that the current sample of the CHIME survey roughly follows this shape with $\alpha = 4$. {However, several surveys are expected to be begin observations in the next years, amongst others the Canadian Hydrogen Observatory and Radio-transient Detector\,(CHORD, \citealt{vanderlinde_2019_chord}) and the Deep Synoptic Array 2000\,(DSA-2000, \citealt{hallinan_2019_dsa}). The latter one is expected to commence survey in March 2026 with an annual localised FRB detection rate of ~10$^4$ and within 3.5\,arcsec resolution, yielding a substantially larger detection rate and increased survey depth in comparison to CHIME, see \citet{connor_2023_stellar} for a more comprehensive review. We thus expect $\alpha=2.5$ to be more representative of the actual redshift distribution feasible within the next years.} We split the redshift distribution into four equipopulated redshift bins (see also \citealt{tanidis_2019_developing}), thus recovering depths information and increasing the signal-to-noise ratio (SNR). We estimate the SNR from the Fisher information matrix \citep{tegmark_karhunen-loeve_1997} by treating the signal amplitude $A$ defined by $\hat{\boldsymbol{C}}_\ell=A\hat{\boldsymbol{S}}_\ell+\hat{\boldsymbol{N}}_\ell$ as a cosmological parameter, in which case the cumulative SNR can be calculated as 
\begin{equation}
\begin{split}
	\label{eq:theo_SNR}
    \Sigma^2(\leq \ell_\mathrm{max}) = & \;\sum_{\ell=1}^{\ell_\mathrm{max}}\mathrm{SNR}^2(\ell) \\
    = & \; f_\mathrm{sky} \sum_{\ell=1}^{\ell_\mathrm{max}}\frac{2\ell+1}{2} \mathrm{Tr}\!\left(\hat{\boldsymbol{C}}_\ell^{-1}\hat{\boldsymbol{S}}_\ell^{\phantom{\ell}}\hat{\boldsymbol{C}}_\ell^{-1}\hat{\boldsymbol{S}}_\ell^{\phantom{\ell}}\right) \;,
    \end{split}
\end{equation}
where $\hat{\boldsymbol{S}}_\ell$, $\hat{\boldsymbol{C}}_\ell$ are the fiducial angular power spectrum matrices without noise and with noise, respectively, and $\ell_\mathrm{max}$ denotes the maximal measured multipole. In Fig. \ref{fig:data_SNR_of_FRB_tomo_z} the differential SNR for different numbers of tomographic redshift bins is depicted. For $n_\mathrm{tomo} > 4$ the gain saturates, hence we will use this as our baseline. One can also see that the SNR peaks around $\ell\sim 200$ due to the relatively large noise contribution which starts dominating at this angular scale.

With the sky coverage factor $f^\mathcal{D}$, the total amount of measured FRBs $n_\mathrm{FRB}$ and the number of tomographic redshift bins $n_\mathrm{tomo}$, the effective number of FRBs per solid angle ($\bar{n}_{\mathrm{FRB},i}$) in radians becomes
\begin{align}
	\bar{n}_{\mathrm{FRB},i} = \frac{n_{\mathrm{FRB}}}{4\pi  f^\mathcal{D}   n_\mathrm{tomo}} \; .
\end{align}
The specifications for the FRB redshift bins are shown in Tab. \ref{tab:Euclid_specs} and the source distribution is depicted in Fig. \ref{fig:FRB_redshift_dist}. 

Discontinuities between the FRB bins originate from the individual normalisation of the source distribution in each bin, such that all $n_i(z)$ are proper probability distribution. Note that we use a simplistic mock FRB catalogue, which implicitly assumes full knowledge of the DM-$z$ relation. In reality, the real redshift of the FRB progenitor is often unknown. That can lead to a multitude of unaccounted for effects in this pipeline, such as the scattering of sources between different redshift bins, similarly to the photometric redshift scatter known from shear analyses\,(\citealt{dahlen_2013_critical}, \citealt{euclid_2020_preparation_X}). While this mock catalogue is a suitable starting point to estimate the performance of future FRB surveys, we refer the reader to future analyses for more elaborate FRB catalogues.

\subsubsection{Shear catalogue}
We adapt a stage-4 WL survey with properties given in \citet{euclid_collaboration_euclid_2020}. The corresponding specifications can be found in Tab. \ref{tab:Euclid_specs} and the source-redshift distribution is shown in Fig. \ref{fig:Euclid_redshift_dist}. We will henceforth refer to the WL survey as Euclid-Cosmic-Shear (\emph{Euclid-CS}) in the remainder of the paper.

\subsection{Three dimensional power spectrum generation}
\label{ch:theory_3dpower_spec}

We calculate the linear matter power spectrum with the Boltzmann solver \textbf{H}orndeski \textbf{i}n \linebreak \texttt{class}\,(\texttt{hi\_class}\footnote{\href{http://miguelzuma.github.io/hi_class_public/}{http://miguelzuma.github.io/hi\_class\_public/}}, \citealp{zumalacarregui_hiclass:_2016}, \citealp{Bellini_2020_hi_class}) suite which is an extension to the \textbf{C}osmic \textbf{L}inear \textbf{A}nisotropy \textbf{S}olving \textbf{S}ystem code (\texttt{class}\footnote{\href{https://github.com/lesgourg/class_public}{https://github.com/lesgourg/class\_public}}, \citealp{blas_cosmic_2011}). Subsequently, utilising \texttt{pyccl}\footnote{\href{https://github.com/LSSTDESC/CCL}{https://github.com/LSSTDESC/CCL}} \citep{pyccl}, we pass the linear matter power spectrum into \texttt{HMcode}\footnote{\href{https://github.com/alexander-mead/HMcode}{https://github.com/alexander-mead/HMcode}} \citep{mead_accurate_2015} to calculate the non-linear matter and electron power spectrum $P_\mathrm{mm}$, $P_\mathrm{\mathrm{ee}}$ and, consequently, the electron bias via
\begin{align}
    b_\mathrm{e}^2(k,z) = \frac{P_\mathrm{ee}(k,z)}{P_\mathrm{mm}(k,z)}\;,
\end{align}
where we employ the halo model designated as \linebreak \texttt{HMx2020\_matter\_pressure\_w\_temp\_scaling} \citep{mead_hydrodynamical_2020}. The full list of cosmological parameters with their respective fiducial values is shown in Tab. 
\ref{tab:data_fid_cosmology}. 

Generating thousands of samples from \texttt{hi\_class} for the MCMC is computationally expensive, in particular if we want to run multiple analyses with varying hyperparameters. Thus, we employ \texttt{cosmopower}\footnote{\href{https://github.com/alessiospuriomancini/cosmopower}{https://github.com/alessiospuriomancini/cosmopower}} \citep{spuriomancini_cosmopower_2022} to train and emulate the nonlinear matter power spectrum $P_\mathrm{mm}$, the electron bias $b_\mathrm{e}$, the comoving distance $\chi$ and the modified gravity functions $\mu,\eta$. The latter two are calculated using a modified \texttt{hi\_class} code from \citet{spurio_mancini_testing_2018}. By initialising \texttt{hi\_class} with the parameters and prior bounds from Tab.\;\ref{tab:data_fid_cosmology}, \textsb{we simulated $147\,410$ training and $9\,992$ testing sets in a Latin-Hyper-Cube}, {meaning they are semi-randomly evenly distributed across the whole hyperspace.} Note that we are limited to a one-dimensional output layer due to the nature of neural networks, hence we treat the redshift dependency of the individual functions as a cosmological parameter with bounds $z\in[0,\,4.5]$ the emulator needs to be trained on. In total, we map the (10+1)-parameter input space to a 200 parameter space consisting of $k$-values logarithmically spaced in $k\in[10^{-5}, 150]\,\mathrm{Mpc}^{-1}$. The highly non-linear upper bound is a numerical necessity caused by the limber approximation since 
\begin{align*}
    k_\mathrm{max} = \frac{\ell_\mathrm{max}+1/2}{\chi(z_\mathrm{min}=10^{-2})} \approx 113\,\mathrm{Mpc}^{-1}
\end{align*}
in Planck cosmology. The majority of signal resides around $\ell\sim\mathcal{O}(200)$ (see also Fig. \ref{fig:data_SNR_of_FRB_tomo_z}) for FRBs. While cosmic shear obtains a significant signal from larger $\ell$, contributions for $k> 50\;\mathrm{Mpc}$ will generally be small. For more details on the emulator choices we refer to appendix \ref{ch:app_accuracy_emulator}.

\begin{figure*}
    \centering
    \includegraphics[width=1\linewidth]{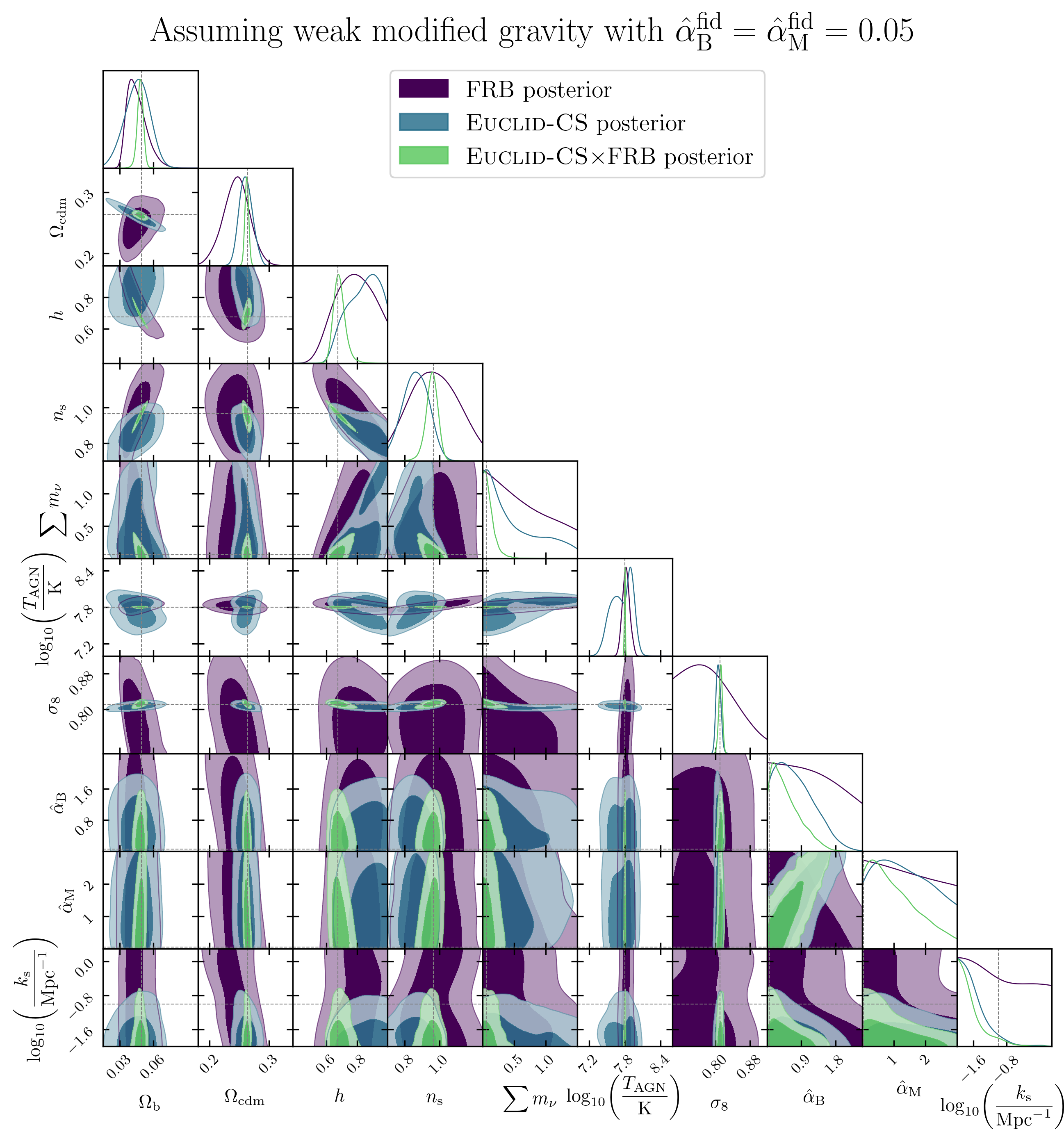}
    \caption{68\% and 95 \% posterior confidence regions (dark and light areas, respectively) for FRB and \textsc{Euclid-CS} auto-correlation and their cross-correlation. The fiducial cosmology taken from Tab. \ref{tab:data_fid_cosmology} is indicated by grey, dashed lines.} 
    \label{fig:res_mcmc_full_contours}
\end{figure*}

Finally, with the emulators we can calculate the angular power spectra detailed in Section\;\ref{ch:theory}, which in turn enables us to obtain the log-likelihood (Eq.\;\ref{eq:data_log_likelihood}) for any cosmology contained in the prior ranges from Tab. \ref{tab:data_fid_cosmology}. Hence, by sampling the likelihood with the Importance Nested Sampling Algorithm\,(INS, \citealt{Feroz2019Importance}) using \texttt{Nautilus}-sampler\footnote{\href{https://github.com/johannesulf/nautilus.git}{https://github.com/johannesulf/nautilus.git}} \citep{lange_nautilus_2023}, we obtain the posterior probability density.

\begin{figure*}
    \centering
    \includegraphics[width=1\linewidth]{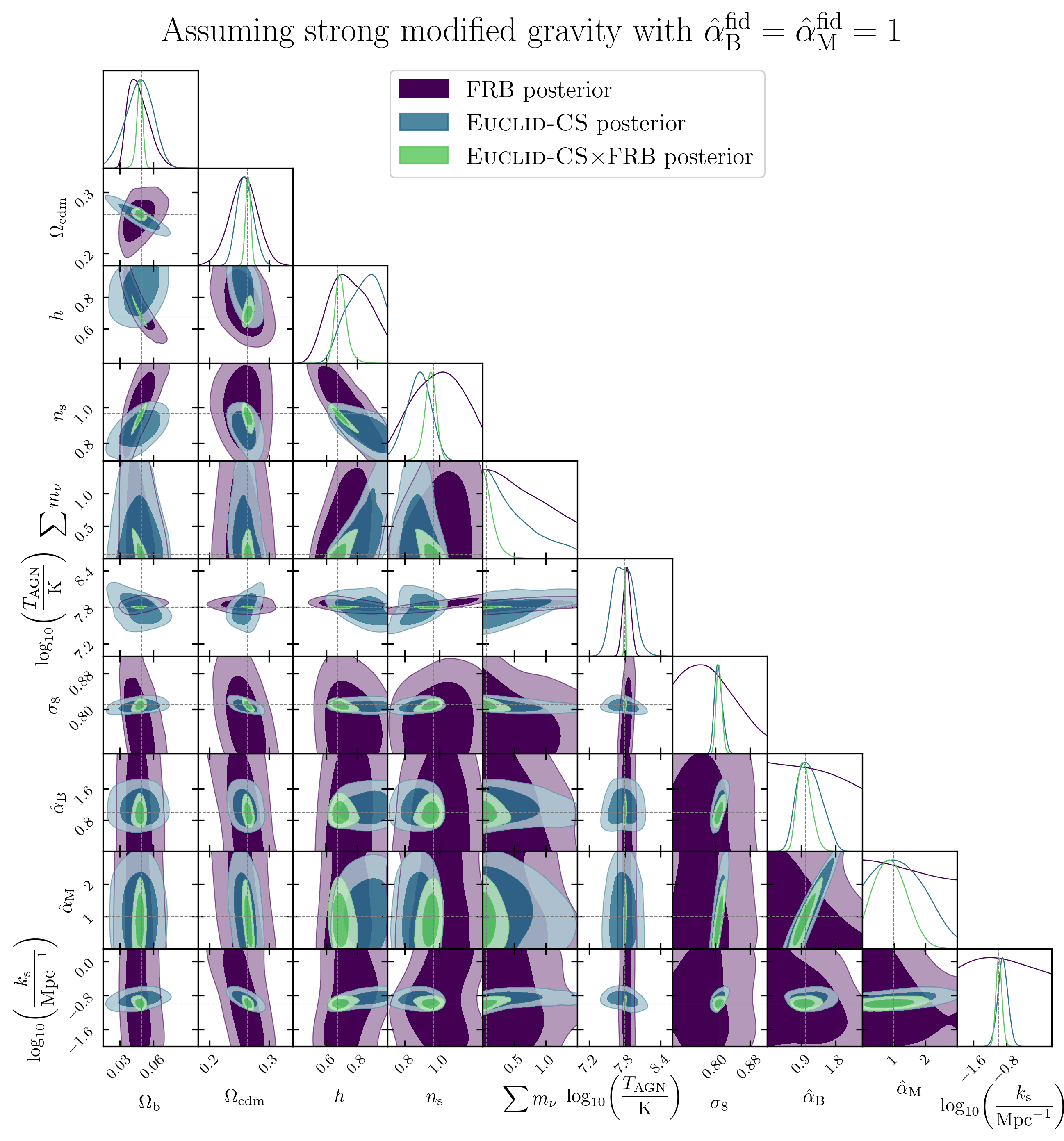}
    \caption{68\% and 95 \% posterior confidence regions (dark and light areas, respectively) for FRB and \textsc{Euclid-CS} auto-correlation and their cross-correlation. The fiducial cosmology from Tab. \ref{tab:data_fid_cosmology} is used except for $\hat{\alpha}_\mathrm{M}=\hat{\alpha}_\mathrm{B}=1$.}
    \label{fig:res_mcmc_full_contours_alpha1}
\end{figure*}

\section{Results}
\label{ch:results}
We show the 68\% and 95\% confidence regions for the three cases, i.e. WL and FRBs alone as well as WL $\times$ FRBs in Fig. \ref{fig:res_mcmc_full_contours}. Overall, FRBs have \linebreak considerably less constraining power than WL alone due to the overall lesser SNR, as WL from a stage-4 survey have a cumulative SNR which is roughly 10 times higher than the one of the FRB mock survey assumed here. Consequently, also the constraints on modified gravity are generally weaker. However, FRBs impose much smaller uncertainties on the baryon density $\Omega_\mathrm{b}h^2$ and the baryonic feedback strength $\Tagn$. The latter has been extensively discussed in our previous analysis \citep{reischke_calibrating_2023}, hence we will refrain from detailing this effect here. Nevertheless, it should be noted that we expect a majority of the cross-correlation improvement to originate from degeneracy breaking due to baryonic feedback calibration with FRBs. Indeed, when cross-correlating shear and DM, the confidence regions of almost all parameters shrink significantly. The marginalised posterior mean and the 68\% confidence intervals are quantified in Tab. \ref{tab:res_alpha_0.05}. After $\Tagn$, the parameters describing the energy contents of the Universe ($\sum m_\nu$, $\Omega_\mathrm{b}$, $\Omega_\mathrm{cdm}$) experience the greatest increase in precision. Looking at the $\Omega_\mathrm{b}$-$h$ degeneracy apparent in Fig. \ref{fig:res_mcmc_full_contours}, a separate calibration of $h$ further tightens the obtained constraints. In appendix \ref{ch:app_adding_planck}, we show the posterior for combining \textsc{Euclid-CS}$\times$FRB with the \textsc{Planck}2018\,\citep{planck_2020_overview} temperature anisotropy auto-correlation. This breaks the degeneracy $\Omega_\mathrm{b}$-$h$, leading to much tighter constraints. \citet{Walters_2018_future} similarly forecasted competitive constraining power when combining FRBs with distance ladder measurement on the background level. Since FRBs can be considered a late-type cosmological probe, this is a sensible addition for future analyses. 
A smaller but still considerable amount of gain is seen for the primordial power spectrum scale dependency $n_\mathrm{s}$, possibly caused by the comparatively large signal of FRBs for small $k$, at which $n_\mathrm{s}$ still determines the slope. 

On the contrary, little information can be drawn for $\sigma_8$ in contrast to the WL measurement, \textsb{leading to only a $\sim24\%$} improvement for the cross-correlation. The improvement of $\hat{\alpha}_\mathrm{M/B}$ is on a similar order of magnitude, putting them amongst the most weakly constrained parameters. \textsb{Nevertheless, a $\sim$$25-35\%$ reduction is detectable}, likely originating from degeneracy breaking between other cosmological parameters and general accessibility of more large scale modes. However, most notably, the $\ks$ posterior has a peculiar behaviour. Including FRBs into the analysis has almost no discernible effect on the precision of this measurement. Additionally, we forecast a bias of the posterior towards low screening scales $k_\mathrm{s}$. WL in particular tends to underpredict this parameter. This is due to projection effects of the full posterior when showing marginalised contours. As the modified gravity parameters are bounded from below by $\Lambda$CDM there is a bunching up of probability in this region when the fiducial (i.e. the real Universe) is very close to $\Lambda$CDM. Since $\ks$ has no effect in $\Lambda$CDM, the posterior is just the prior in this case. Therefore, the projection to a marginalised contour can appear to bias the parameters. Increasing the prior range to lower bounds would thus not increase the gain in information, apart from removing almost any modified gravity effects altogether.

Further peculiar behaviour can be seen for the $\Tagn$-\textsc{Euclid-CS} posterior in Fig. \ref{fig:res_mcmc_full_contours}; displaying a bimodal distribution. While a Gaussian probability distribution cannot be expected in an MCMC, specifically the modified gravity parameter contour shapes show large deviations thereof caused by limiting prior ranges. We excluded significant influence from the MCMC-algorithm as the underlying cause (see App.\;\ref{ch:app_convergence_test}). Considering the strong deviation of the predicted $k_\mathrm{s}$ to the fiducial one, we conclude that the contours from Fig. \ref{fig:res_mcmc_full_contours} to be strongly influenced by the informative prior we place on $\hat{\alpha}_{\mathrm{B/M}}$ and $k_\mathrm{s}$.

The motivation behind this choice is of both physical and numerical nature. In principle, $\hat{\alpha}_{\mathrm{B/M}}$ can be negative \citep{bellini_maximal_2014} and even are expected to in many of the commonly investigated scalar-field theories like $f(R)$ \citep{carroll_2004_cosmic} or Galileon cosmology \citep{chow_2009_galileon}. However, $\hat{\alpha}_{\mathrm{B/M}}<0$ can lead to gradient instabilities produced by an imaginary sound speed that causes exponentially growing perturbations \citep{zumalacarregui_hiclass:_2016}. Even on the theoretically stable $\hat{\alpha}_\mathrm{M}=-\hat{\alpha}_\mathrm{B}$ line, numerical fluctuations can lead to the aforementioned runaway effect. We demonstrate this effect by showing the $\hat{\alpha}_\mathrm{M}-\hat{\alpha}_\mathrm{B}$ parameter-space with successfully initialised \texttt{hi\_class} instances in \linebreak Fig. \ref{fig:res_aBM_exploration}. Note that we simulate within the prior ranges from Tab.\;\ref{tab:data_fid_cosmology} except for the braiding and Planck mass run rate, which we extend to $\hat{\alpha}_\mathrm{B/M}\in[-3,3]$. Thus, while certain sets of negative parameters are stable \citep{spurio_mancini_kids_2019}, uniform and independent priors for $\hat{\alpha}_{\mathrm{B}}$ and $\hat{\alpha}_\mathrm{M}$ are only possible for positive values thereof, hence the choice in this work. {Additionally, we exclude $\aB>2.5$ due to a decreased density of successful \texttt{hi\_class} initialisations there as well.}

\begin{figure}
    \centering
    \includegraphics[width = .45\textwidth]{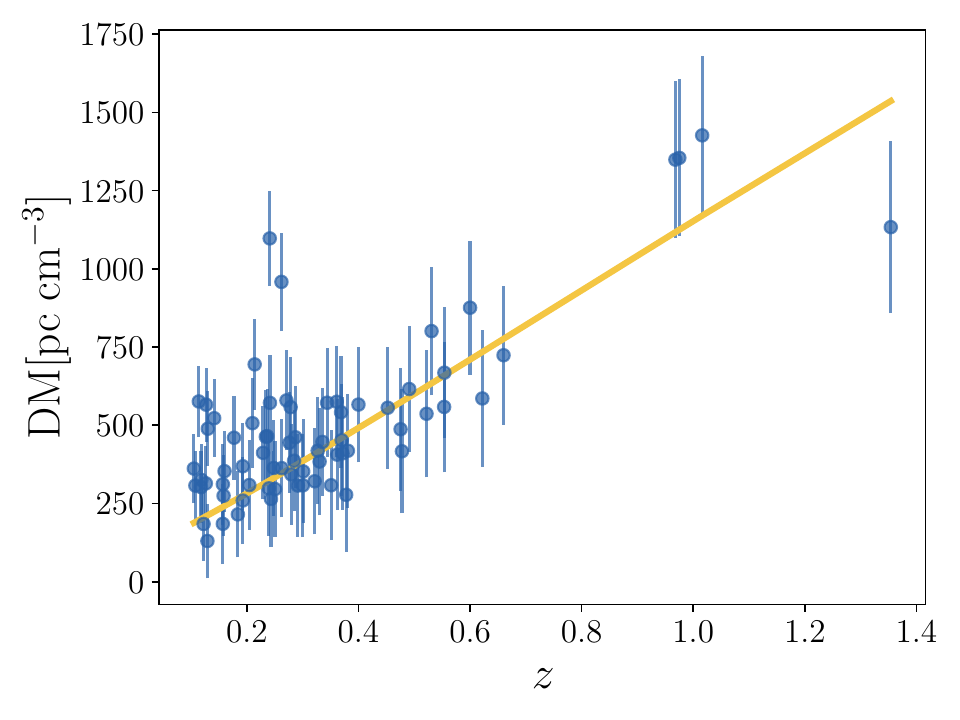}
    \caption{FRB with host identification used in this study. The error bars are a sum of $\sigma_\mathrm{MW}= 50$, $\sigma_\mathrm{host}= 50/(1+z)$ and the cosmological contribution from Eq.\;\ref{eq:cov_gauss}. The orange line corresponds to the best-fit model.}
    \label{fig:model_host}
\vspace{.25cm}\end{figure}

\begin{figure}
    \centering
    \includegraphics[width = .45\textwidth]{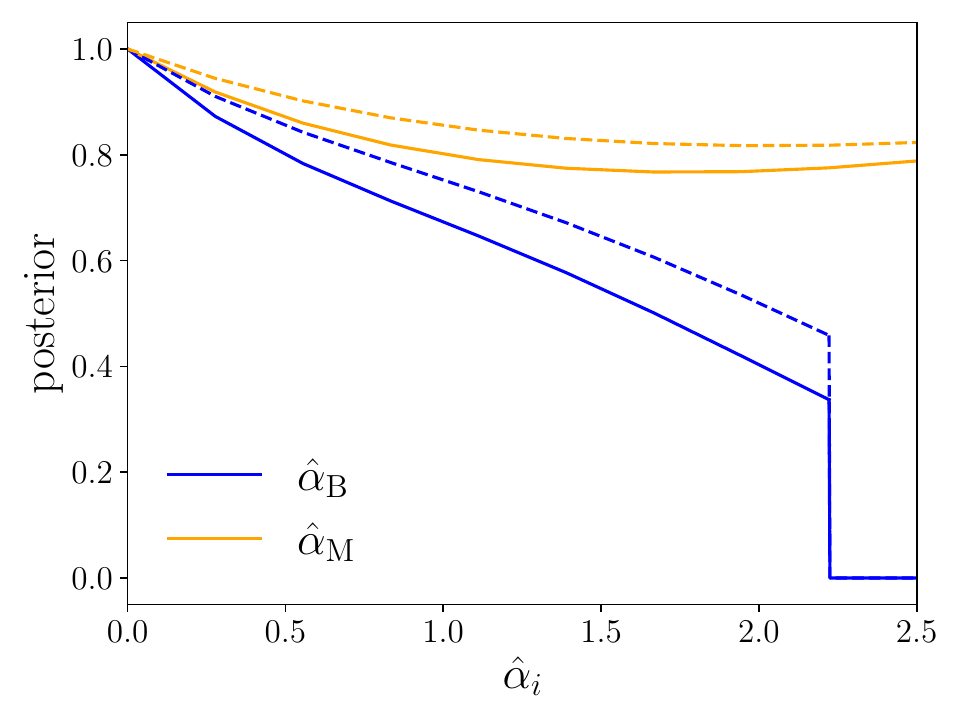}
    \caption{Posterior distribution of $\hat{\alpha}_\mathrm{B}$ (blue) and $\hat{\alpha}_\mathrm{M}$ (orange) assuming a Gaussian (lognormal) host model in solid (dashed). {The cutoff of $\aB$ originates from prior limitations due to \texttt{hi\_class} instabilities (see also Fig.\;\ref{fig:res_aBM_exploration}).}}
    \label{fig:fit_host}
\vspace{.25cm}\end{figure}

To circumvent the issues of those projection affects, we additionally run our analysis for a ``strong fiducial modified gravity'' ($\hat{\alpha}_\mathrm{B/M}^\mathrm{fid}=1$) as opposed to the previously used ``weak fiducial modified gravity'' ($\hat{\alpha}_\mathrm{B/M}^\mathrm{fid}=0.05$). We know from the \textsc{Planck} CMB survey \citep{planck_collaboration_xiv_planck_2015} that we live at least in a close-to-$\Lambda$CDM Universe. Still, by increasing the fiducial braiding and Planck mass run rate parameters, we move them away from the prior edges and, hence, mimic looser boundaries. We do not quote the marginalised posterior uncertainties explicitly, but show the improvement when using FRBs in brackets in the last column of Tab. \ref{tab:res_alpha_0.05}. However, we show the contours in Fig. \ref{fig:res_mcmc_full_contours_alpha1}. First, one should notice that much of the peculiar behaviour of the WL contours vanish in strong modified gravity; the WL $\Tagn$ posterior is no longer bimodal and the measurements of $k_\mathrm{s}$ are not biased towards lower values anymore. The screening scale recovered now is on par with the fiducial input screening scale, chosen such as to suppress modified gravity in the non-linear regime (see also \citealt{spurio_mancini_testing_2018}).
 
Consequently, the 68\% confidence intervals change slightly as well. Even though FRBs still can not constrain modified gravity alone, \textsb{the improvement of $\hat{\alpha}_\mathrm{B}$ from WL auto-correlation to WL-FRB cross-correlation slightly increases to $42.5\%$ and the one of $\hat{\alpha}_\mathrm{M}$ to $34.6\%$}. Other parameter improvements are comparable to the weak modified gravity case, \textsb{with the slight exception of $\sum m_\nu$ and $\Omega_{\rm cdm}$ where the improvement decreases in strong modified gravity, and $\Omega_{\rm b}$ where it increases}. Therefore, we expect to gain constraining power on modified gravity when cross-correlating FRBs with WL, albeit less sizeable than for baryon-related parameters. {Adding CMB temperature anisotropy information has little effect on the modified gravity parameter contours, indicating only weak correlation between $\hat{\alpha}_\mathrm{B/M}$ and $h$, $\Omega_\mathrm{b}$, $n_\mathrm{s}$ (see appendix \ref{ch:app_adding_planck}).}

\section{Application to FRBs with host}
\label{ch:appli}

In this section, we use FRBs with host identification to provide some upper bounds on modified gravity. This is very similar to the methodology outlined in \citet{reischke_cosmological_2023,2023_reischke_consistent} where the sensitivity is in the covariance between the different data points of the measurement of the $\mathrm{DM}$-$z$ relation, i.e. probing correlations between different lines-of-sight. There are already some analyses of statistical properties of the signal via cross-correlation with the LSS \citep[e.g.][]{chime_2021_first} or using targeted galaxy surveys to reduce cosmic variance \citep{khrykin_flimflam_2024}. However, the quality of the current data is not good enough yet to carry out the analysis proposed in the previous section. 
The LSS covariance of two FRBs observed at redshift $z_i, z_j$ and with pairwise separation $ {\theta}_{ij}$ can be calculated as 
\begin{equation}
\label{eq:cov_gauss}
    \mathrm{cov}_{ij}(\cos \theta, z_i, z_j) = \sum_\ell \frac{2\ell+1}{4\pi}P_\ell(\cos \theta)\left(C^\mathcal{DD}_\ell\right)_{ij}\;,
\end{equation}
where we use a $\delta$-distribution for the redshift distribution in Eq.\;\ref{eq:theo_DM_kernel} selecting redshift $z_i$ and $z_j$. Here the Legendre polynomials are denoted as $P_\ell(x)$.

Assuming a distribution for the host, $p_\mathrm{host}$, the final likelihood is given by:
    \begin{equation}
    \begin{split}
        p(\boldsymbol{\mathrm{DM}}|\boldsymbol{\theta}) = &\;\int_0^\infty\mathrm{dDM}_\mathrm{host}\;p_\mathrm{host}( \boldsymbol{\mathrm{DM}}_\mathrm{host})\\ &\times \;p_\mathrm{LSS}(\boldsymbol{\mathrm{DM}} - \boldsymbol{\mathrm{DM}}_\mathrm{host})\;,    
    \end{split}
    \end{equation}
    where $ (\boldsymbol{\mathrm{DM}}_\mathrm{host})_i =  \mathrm{DM}_\mathrm{host}/(1+z_i)$ and $p_\mathrm{LSS}$ is a Gaussian with covariance with components given by Eq.\;\ref{eq:cov_gauss} and the components of the mean by:    
\begin{equation}
\label{eq:DM_LSS_avg}
    \mathrm{DM}_\mathrm{LSS}(z_i)  = \int_0^{\chi(z_i)} \mathrm{d}\chi^\prime W^{(i)}_\mathcal{D}(\chi^\prime)\;.
\end{equation}
In Fig. \ref{fig:model_host} we show the FRBs used for this analysis together with the best fit model, {see Tab.\;\ref{tab:app_frb_list} in the Appendix for a comprehensive list}. The error includes the covariance contribution from the LSS, from the host and from the Milky Way. Due to the low number of FRBs, we maximize the likelihood with respect to the $\Lambda$CDM parameters and then scan the profiles of $\hat{\alpha}_\mathrm{M}$ and $\hat{\alpha}_\mathrm{B}$ respectively. This removes the issue of projection effects and allows to get the most stringent constraints from the current data set.
Fig. \ref{fig:fit_host} shows this likelihood scan, showing that the posterior is prior driven (see Tab. \ref{tab:data_fid_cosmology}). Note that the sharp cut-off in $\aB$ is caused by the introduced emulator boundaries. There seems to be no information regarding $\hat{\alpha}_\mathrm{M}$ with the current data. While there is a slight constraint on $\hat{\alpha}_\mathrm{B}$, this is not significantly different from the prior. Generally this behaviour is expected as the parameter dependence is only present in the perturbations and thus in the covariance, degrading the information content compared to the model. However, it goes to show that host-identified FRBs are in principle a versatile tool to test gravity, in particular with many more sources expected in the next years \citep[e.g. with DSA-2000,][]{hallinan_2019_dsa}

\section{Summary}
\label{ch:conclusion}
In this paper, we investigated the prospects to constrain modified gravity with the DM of FRB and WL, as well as their cross-correlation. We calculated the linear Horndeski-type perturbations of GR (i.e. a metric theory with an additional scalar degree of freedom) that acts on the bardeen potentials $\Phi$ and $\Psi$. Non-linear correction to that model has been applied with $\Lambda$CDM corrections using \texttt{HMcode} with baryonic feedback controlled by a linear interpolation of a one-parametric model.
Since FRBs show a substantial signal-to-noise ratio for small $k$-scales while being very sensitive to baryonic effects, they complement WL. We employed the Boltzmann-solver \texttt{hi\_class} to simulate the background cosmology and linear perturbations thereof governed by modified gravity. By propagating the results through \texttt{HMcode}, we obtain the non-linear corrections to the three-dimensional matter and electron power spectra in a $\Lambda$CDM cosmology. We ensure screening by introducing a free parameter $k_\mathrm{s}$ so that GR is recovered for $k > k_\mathrm{s}$. The three-dimensional power spectra are then projected onto angular space with the specified \textsc{Euclid-CS} and 10000 sources-FRB survey. This 10-parameter model is sensitive to modified gravity on linear scales, baryonic effects and matter clustering and can be a starting point for future WL$\times$FRB large-scale structure analyses. To speed up calculations, we trained a \texttt{cosmopower} neural network to emulate the statistical properties of the Bardeen potentials and the matter and the electron power spectrum.
We carried out three types of analysis:
\begin{enumerate}
    \item[i)] First, we investigated a weak modified gravity case with fiducial linear Planck mass run rate and braiding proportionality constant set to $\aM^{\mathrm{fid}}=\aB^{\mathrm{fid}}=0.05$.  In this case, the forecast is dominated by the $\hat{\alpha}_\mathrm{M/B}$ prior edges that exclude negative values. We additionally detect a preference for low values of the screening scale $k_\mathrm{s}$. However, the posterior is generally rather unconstrained, which we believe to be a consequence of the fiducial cosmology at which the data was drawn. Nevertheless, by combining 10000 FRBs with redshift information with a \textsc{Euclid-CS}-like survey, \textsb{the $\aM$ and $\aB$ 68\%-constraints improve by a factor of 1.3 and 1.5 respectively, in comparison to the constraints from \textsc{Euclid-CS} alone.}

    \item[ii)]  We repeat the same analysis with strong fiducial modified gravity case with $\aM^{\mathrm{fid}}=\aB^{\mathrm{fid}}=1$. Here the posteriors are no longer prior dominated and $k_\mathrm{s}$-bias is no longer detectable. The improvement from WL to WL$\times$FRB \textsb{changes to a factor of 1.7, 1.5 and 1.7 for $\aM$ $\aB$ and $\ks$ respectively.} FRBs alone put little constraints on Horndeski theory, so the majority of signal in the cross-correlation with WL likely originates from degeneracy breaking between other cosmological parameters. It has already been established that FRBs are highly sensitive to the intergalactic baryonic component, which is where a majority of the signal resides in this work as well. \textsb{Nevertheless, a 40\% decrease in the error bars makes an implementation of FRBs into future modified gravity investigations worthwhile.}

    \item[iii)] \textsb{We conduct an analysis of 71 FRBs with host identification} where the dependence of modified gravity is expressed in the covariance of the $\mathrm{DM}-z$ relation of the FRBs. The constraints on modified gravity are prior dominated in this case with a small bit of information added for $\aB$, consistent with the other findings.
    
\end{enumerate}
During the weak modified gravity analysis, core limitations of the underlying modelling became apparent. Even though many scalar-tensor perturbations of GR require a negative scaling of the braiding or Planck mass run rate with dark energy, this parameter space is largely unstable, and thus uncontainable, in the current \texttt{hi\_class} release. This is in large part due to the obvious necessity of non-imaginary primordial baryon sound velocity. That being said, because of numerical fluctuations the $\aM=-\aB$ line is non-accessible as well, despite the fact that is physically conceivable. While we can derive some limits as demonstrated above, the current state of general, numerical Horndeski theory tools is not fit for large-scale cosmological analyses, yet. Possible refinements for \texttt{hi\_class} could include shifting the parameterisation from $\aM$ or $\aB$ to the sound velocity $c_s^2$ such that small numerical deviations can not lead to a runaway gradient. In a recent development, \citet{cataneo_2024_mochi_class} built the \emph{Modelling Optimisation to Compute Horndeski in CLASS}\,(\texttt{mochi\_class}\footnote{\href{https://github.com/mcataneo/mochi_class_public}{https://github.com/mcataneo/mochi\_class\_public}}) suite, which might be able to address some of these issues in future analyses.

Generally, we find FRBs to be a versatile assisting tool for WL to constrain the nature of gravity. While they are not very sensitive to modifications to gravity by themselves, they allow to break degeneracies in other parameters and constrain baryonic feedback \citep{reischke_calibrating_2023}, thus freeing up more WL signal to be used for the measurement of the gravitational slip. FRBs are thus a suitable addition to other LSS probes like WL that will soon likely gain enough precision to confirm or rule out such scalar perturbations, causing the necessity of accurate modified gravity modelling tools.   

\section*{Acknowledgements}
During parts of this work, RR was supported by a European Research Council Consolidator Grant (No. 770935). DN acknowledges funding from the European Research Council (ERC) under the European Union's Horizon 2020 research and innovation program (Grant agreement No. 101053992). SH was supported by the Excellence Cluster ORIGINS which is funded by the Deutsche Forschungsgemeinschaft (DFG, German Research Foundation) under Germany’s Excellence Strategy - EXC-2094 - 390783311. RR is supported by the ERC (Grant No. 770935). H. Hildebrandt is supported by a DFG Heisenberg grant (Hi 1495/5-1), the DFG Collaborative Research Center SFB1491, an ERC Consolidator Grant (No. 770935), and the DLR project 50QE2305.

\bibliographystyle{mnras}
\bibliography{library1,library2}

\begin{thebibliography}{}
\makeatletter
\relax
\def\mn@urlcharsother{\let\do\@makeother \do\$\do\&\do\#\do\^\do\_\do\%\do\~}
\def\mn@doi{\begingroup\mn@urlcharsother \@ifnextchar [ {\mn@doi@} {\mn@doi@[]}}
\def\mn@doi@[#1]#2{\def\@tempa{#1}\ifx\@tempa\@empty \href {http://dx.doi.org/#2} {doi:#2}\else \href {http://dx.doi.org/#2} {#1}\fi \endgroup}
\def\mn@eprint#1#2{\mn@eprint@#1:#2::\@nil}
\def\mn@eprint@arXiv#1{\href {http://arxiv.org/abs/#1} {{\tt arXiv:#1}}}
\def\mn@eprint@dblp#1{\href {http://dblp.uni-trier.de/rec/bibtex/#1.xml} {dblp:#1}}
\def\mn@eprint@#1:#2:#3:#4\@nil{\def\@tempa {#1}\def\@tempb {#2}\def\@tempc {#3}\ifx \@tempc \@empty \let \@tempc \@tempb \let \@tempb \@tempa \fi \ifx \@tempb \@empty \def\@tempb {arXiv}\fi \@ifundefined {mn@eprint@\@tempb}{\@tempb:\@tempc}{\expandafter \expandafter \csname mn@eprint@\@tempb\endcsname \expandafter{\@tempc}}}

\bibitem[\protect\citeauthoryear{{Abbott}}{{Abbott}}{2024}]{abbott_chimefrb_20240202}
{Abbott} T.,  2024, TNS FRBs, \href {https://ui.adsabs.harvard.edu/abs/2024TNSFR.330....1A} {330, 1}

\bibitem[\protect\citeauthoryear{Abbott et~al.,}{Abbott et~al.}{2017}]{abbott_gravitational_2017}
Abbott B.~P.,  et~al., 2017, \mn@doi [ApJ] {10.3847/2041-8213/aa920c}, 848, L13

\bibitem[\protect\citeauthoryear{Adi \& Kovetz}{Adi \& Kovetz}{2021}]{adi_2021_probing}
Adi T.,  Kovetz E.~D.,  2021, \mn@doi [\prd] {10.1103/PhysRevD.104.103515}, 104, 103515

\bibitem[\protect\citeauthoryear{Aghanim}{Aghanim}{2001}]{aghanim_2001_reionisation}
Aghanim N.,  2001, \mn@doi [\nar] {https://doi.org/10.1016/S1387-6473(00)00152-4}, 45, 303

\bibitem[\protect\citeauthoryear{Alonso, Bellini, Ferreira  \& Zumalac\'arregui}{Alonso et~al.}{2017}]{alonso_2017_observational}
Alonso D.,  Bellini E.,  Ferreira P.~G.,   Zumalac\'arregui M.,  2017, \mn@doi [\prd] {10.1103/PhysRevD.95.063502}, 95, 063502

\bibitem[\protect\citeauthoryear{Alonso, Sanchez, Slosar  \& {LSST Dark Energy Science Collaboration}}{Alonso et~al.}{2019}]{alonso_unified_2019}
Alonso D.,  Sanchez J.,  Slosar A.,   {LSST Dark Energy Science Collaboration} 2019, \mn@doi [\mnras] {10.1093/mnras/stz093}, 484, 4127

\bibitem[\protect\citeauthoryear{{Arcus}, {Macquart}, {Sammons}, {James}  \& {Ekers}}{{Arcus} et~al.}{2021}]{arcus_2021_fast}
{Arcus} W.~R.,  {Macquart} J.~P.,  {Sammons} M.~W.,  {James} C.~W.,   {Ekers} R.~D.,  2021, \mn@doi [\mnras] {10.1093/mnras/staa3948}, 501, 5319

\bibitem[\protect\citeauthoryear{Auclair et~al.,}{Auclair et~al.}{2023}]{Auclair_2023_LISA}
Auclair P.,  et~al., 2023, \mn@doi [Living Rev. Relativ.] {10.1007/s41114-023-00045-2}, 26, 5

\bibitem[\protect\citeauthoryear{Babichev \& Deffayet}{Babichev \& Deffayet}{2013}]{babichev_introduction_2013}
Babichev E.,  Deffayet C.,  2013, \mn@doi [Classical Quant. Grav.] {10.1088/0264-9381/30/18/184001}, 30, 184001

\bibitem[\protect\citeauthoryear{Bannister et~al.,}{Bannister et~al.}{2019}]{bannister_single_2019}
Bannister K.~W.,  et~al., 2019, \mn@doi [Science] {10.1126/science.aaw5903}, 365, 565

\bibitem[\protect\citeauthoryear{Bardeen}{Bardeen}{1980}]{bardeen_gauge-invariant_1980}
Bardeen J.~M.,  1980, \mn@doi [\prd] {10.1103/PhysRevD.22.1882}, 22, 1882

\bibitem[\protect\citeauthoryear{Bellini \& Sawicki}{Bellini \& Sawicki}{2014}]{bellini_maximal_2014}
Bellini E.,  Sawicki I.,  2014, \mn@doi [\jcap] {10.1088/1475-7516/2014/07/050}, 7, 050

\bibitem[\protect\citeauthoryear{Bellini, Cuesta, Jimenez  \& Verde}{Bellini et~al.}{2016}]{bellini_constraints_2016}
Bellini E.,  Cuesta A.~J.,  Jimenez R.,   Verde L.,  2016, \mn@doi [\jcap] {10.1088/1475-7516/2016/06/E01}, 02, 053

\bibitem[\protect\citeauthoryear{Bellini, Sawicki  \& Zumalacárregui}{Bellini et~al.}{2020}]{Bellini_2020_hi_class}
Bellini E.,  Sawicki I.,   Zumalacárregui M.,  2020, \mn@doi [\jcap] {10.1088/1475-7516/2020/02/008}, 2020, 008

\bibitem[\protect\citeauthoryear{{Bhandari}, {Kumar}, {Shannon}  \& {Macquart}}{{Bhandari} et~al.}{2019}]{shannon_2019_ASKAP190714}
{Bhandari} S.,  {Kumar} P.,  {Shannon} R.~M.,   {Macquart} J.~P.,  2019, The Astronomer's Telegram, \href {https://ui.adsabs.harvard.edu/abs/2019ATel12940....1B} {12940, 1}

\bibitem[\protect\citeauthoryear{Bhandari et~al.,}{Bhandari et~al.}{2022}]{bhandari_characterizing_2022}
Bhandari S.,  et~al., 2022, \mn@doi [\aj] {10.3847/1538-3881/ac3aec}, 163, 69

\bibitem[\protect\citeauthoryear{{Bhattacharya}, {Kumar}  \& {Linder}}{{Bhattacharya} et~al.}{2021}]{bhattacharya_fast_2020}
{Bhattacharya} M.,  {Kumar} P.,   {Linder} E.~V.,  2021, \mn@doi [\prd] {10.1103/PhysRevD.103.103526}, \href {https://ui.adsabs.harvard.edu/abs/2021PhRvD.103j3526B} {103, 103526}

\bibitem[\protect\citeauthoryear{Blanchard et~al.}{Blanchard et~al.}{2020}]{blanchard_euclid_2020}
Blanchard A.,  et~al., 2020, \mn@doi [Astron. Astrophys.] {10.1051/0004-6361/202038071}, 642, A191

\bibitem[\protect\citeauthoryear{Blas, Lesgourgues  \& Tram}{Blas et~al.}{2011}]{blas_cosmic_2011}
Blas D.,  Lesgourgues J.,   Tram T.,  2011, \mn@doi [\jcap] {10.1088/1475-7516/2011/07/034}, 07, 034

\bibitem[\protect\citeauthoryear{Bose, Tsedrik, Kennedy, Lombriser, Pourtsidou  \& Taylor}{Bose et~al.}{2023}]{bose_2023_fast}
Bose B.,  Tsedrik M.,  Kennedy J.,  Lombriser L.,  Pourtsidou A.,   Taylor A.,  2023, \mn@doi [\mnras] {10.1093/mnras/stac3783}, 519, 4780

\bibitem[\protect\citeauthoryear{{CHIME/FRB Collaboration} et~al.,}{{CHIME/FRB Collaboration} et~al.}{2021}]{chime_2021_first}
{CHIME/FRB Collaboration} et~al., 2021, \mn@doi [\apjs] {10.3847/1538-4365/ac33ab}, 257, 59

\bibitem[\protect\citeauthoryear{Caleb et~al.,}{Caleb et~al.}{2023}]{caleb_2023_subarcsec}
Caleb M.,  et~al., 2023, \mn@doi [\mnras] {10.1093/mnras/stad1839}, 524, 2064

\bibitem[\protect\citeauthoryear{Carroll, Duvvuri, Trodden  \& Turner}{Carroll et~al.}{2004}]{carroll_2004_cosmic}
Carroll S.~M.,  Duvvuri V.,  Trodden M.,   Turner M.~S.,  2004, \mn@doi [Phys. Rev. D] {10.1103/PhysRevD.70.043528}, 70, 043528

\bibitem[\protect\citeauthoryear{Cataneo \& Bellini}{Cataneo \& Bellini}{2024}]{cataneo_2024_mochi_class}
Cataneo M.,  Bellini E.,  2024, \mn@doi [Open J. Astroph.] {10.33232/001c.123470}, 7

\bibitem[\protect\citeauthoryear{Cataneo, Lombriser, Heymans, Mead, Barreira, Bose  \& Li}{Cataneo et~al.}{2019}]{cantaneo_2019_on}
Cataneo M.,  Lombriser L.,  Heymans C.,  Mead A.~J.,  Barreira A.,  Bose S.,   Li B.,  2019, \mn@doi [\mnras] {10.1093/mnras/stz1836}, 488, 2121

\bibitem[\protect\citeauthoryear{Chisari et~al.,}{Chisari et~al.}{2019a}]{chisari_2019_modelling}
Chisari N.~E.,  et~al., 2019a, \mn@doi [Open J. Astroph.] {10.21105/astro.1905.06082}, 2

\bibitem[\protect\citeauthoryear{Chisari et~al.,}{Chisari et~al.}{2019b}]{pyccl}
Chisari N.~E.,  et~al., 2019b, \mn@doi [ApJS] {10.3847/1538-4365/ab1658}, 242, 2

\bibitem[\protect\citeauthoryear{Chow \& Khoury}{Chow \& Khoury}{2009}]{chow_2009_galileon}
Chow N.,  Khoury J.,  2009, \mn@doi [Phys. Rev. D] {10.1103/PhysRevD.80.024037}, 80, 024037

\bibitem[\protect\citeauthoryear{Connor \& Ravi}{Connor \& Ravi}{2023}]{connor_2023_stellar}
Connor L.,  Ravi V.,  2023, \mn@doi [\mnras] {10.1093/mnras/stad667}, 521, 4024

\bibitem[\protect\citeauthoryear{{Cordes} \& {Lazio}}{{Cordes} \& {Lazio}}{2002}]{cordes_new_2002}
{Cordes} J.~M.,  {Lazio} T.~J.~W.,  2002, {NE2001.I. A New Model for the Galactic Distribution of Free Electrons and its Fluctuations}, \mn@doi{10.48550/arXiv.astro-ph/0207156}

\bibitem[\protect\citeauthoryear{{DESI Collaboration} et~al.,}{{DESI Collaboration} et~al.}{2025}]{DESI_2024_cosmological}
{DESI Collaboration} et~al., 2025, \mn@doi [\jcap] {10.1088/1475-7516/2025/02/021}, 2025, 021

\bibitem[\protect\citeauthoryear{Dahlen et~al.,}{Dahlen et~al.}{2013}]{dahlen_2013_critical}
Dahlen T.,  et~al., 2013, \mn@doi [\apj] {10.1088/0004-637X/775/2/93}, 775, 93

\bibitem[\protect\citeauthoryear{{De Rham} \& Melville}{{De Rham} \& Melville}{2018}]{rham_2018_gravitational}
{De Rham} C.,  Melville S.,  2018, \mn@doi [Phys. Rev. Lett.] {10.1103/PhysRevLett.121.221101}, 121, 221101

\bibitem[\protect\citeauthoryear{Dewdney, Hall, Schilizzi  \& Lazio}{Dewdney et~al.}{2009}]{dewdney_2009_square}
Dewdney P.~E.,  Hall P.~J.,  Schilizzi R.~T.,   Lazio T. J. L.~W.,  2009, \mn@doi [Proc. IEEE] {10.1109/JPROC.2009.2021005}, 97, 1482

\bibitem[\protect\citeauthoryear{Escamilla, Giarè, Valentino, Nunes  \& Vagnozzi}{Escamilla et~al.}{2024}]{Escamilla_2024_state}
Escamilla L.~A.,  Giarè W.,  Valentino E.~D.,  Nunes R.~C.,   Vagnozzi S.,  2024, \mn@doi [\jcap] {10.1088/1475-7516/2024/05/091}, 2024, 091

\bibitem[\protect\citeauthoryear{{Euclid Collaboration} et~al.,}{{Euclid Collaboration} et~al.}{2020a}]{euclid_collaboration_euclid_2020}
{Euclid Collaboration} et~al., 2020a, \mn@doi [A\&A] {10.1051/0004-6361/202038071}, 642, A191

\bibitem[\protect\citeauthoryear{{Euclid Collaboration} et~al.,}{{Euclid Collaboration} et~al.}{2020b}]{euclid_2020_preparation_X}
{Euclid Collaboration} et~al., 2020b, \mn@doi [A&A] {10.1051/0004-6361/202039403}, 644, A31

\bibitem[\protect\citeauthoryear{Ezquiaga \& Zumalac\'arregui}{Ezquiaga \& Zumalac\'arregui}{2017}]{ezquiaga_2017_dark}
Ezquiaga J.~M.,  Zumalac\'arregui M.,  2017, \mn@doi [Phys. Rev. Lett.] {10.1103/PhysRevLett.119.251304}, 119, 251304

\bibitem[\protect\citeauthoryear{Feroz, Hobson, Cameron  \& Pettitt}{Feroz et~al.}{2019}]{Feroz2019Importance}
Feroz F.,  Hobson M.~P.,  Cameron E.,   Pettitt A.~N.,  2019, \mn@doi [Open J. Astroph.] {10.21105/astro.1306.2144}, 2

\bibitem[\protect\citeauthoryear{Fixsen}{Fixsen}{2009}]{fixsen_temperature_2009}
Fixsen D.~J.,  2009, \mn@doi [ApJ] {10.1088/0004-637X/707/2/916}, 707, 916

\bibitem[\protect\citeauthoryear{Hagstotz, Reischke  \& Lilow}{Hagstotz et~al.}{2022}]{hagstotz_new_2022}
Hagstotz S.,  Reischke R.,   Lilow R.,  2022, \mn@doi [\mnras] {10.1093/mnras/stac077}, 511, 662

\bibitem[\protect\citeauthoryear{{Hallinan} et~al.,}{{Hallinan} et~al.}{2019}]{hallinan_2019_dsa}
{Hallinan} G.,  et~al., 2019, {The DSA-2000 {\textemdash} A Radio Survey Camera}, \mn@doi{10.48550/arXiv.1907.07648}

\bibitem[\protect\citeauthoryear{Hashimoto et~al.,}{Hashimoto et~al.}{2020}]{hashimoto_2020_fast}
Hashimoto T.,  et~al., 2020, \mn@doi [\mnras] {10.1093/mnras/staa2238}, 497, 4107

\bibitem[\protect\citeauthoryear{Heisenberg}{Heisenberg}{2019}]{Heisenberg_2018_large}
Heisenberg L.,  2019, \mn@doi [Phys. Rep.] {https://doi.org/10.1016/j.physrep.2018.11.006}, 796, 1

\bibitem[\protect\citeauthoryear{Horndeski}{Horndeski}{1974}]{horndeski_second-order_1974}
Horndeski G.~W.,  1974, \mn@doi [Int. J. Theor. Phys.] {10.1007/BF01807638}, 10, 363

\bibitem[\protect\citeauthoryear{James et~al.,}{James et~al.}{2022}]{james_measurement_2022}
James C.~W.,  et~al., 2022, \mn@doi [\mnras] {10.1093/mnras/stac2524}, 516, 4862

\bibitem[\protect\citeauthoryear{Jiang, Ren, Li, Cai  \& Er}{Jiang et~al.}{2024}]{jiang_2024_exploring}
Jiang X.,  Ren X.,  Li Z.,  Cai Y.-F.,   Er X.,  2024, Exploring $f(T)$ Gravity via strongly lensed fast radio bursts, \mn@doi{10.48550/arXiv.2401.05464}

\bibitem[\protect\citeauthoryear{Joyce, Jain, Khoury  \& Trodden}{Joyce et~al.}{2015}]{joyce_beyond_2015}
Joyce A.,  Jain B.,  Khoury J.,   Trodden M.,  2015, \mn@doi [\physrep] {10.1016/j.physrep.2014.12.002}, 568, 1

\bibitem[\protect\citeauthoryear{Khoury \& Weltman}{Khoury \& Weltman}{2004a}]{khoury_2004_chameleon}
Khoury J.,  Weltman A.,  2004a, \mn@doi [Phys. Rev. D] {10.1103/PhysRevD.69.044026}, 69, 044026

\bibitem[\protect\citeauthoryear{Khoury \& Weltman}{Khoury \& Weltman}{2004b}]{khoury_chameleon_2004-1}
Khoury J.,  Weltman A.,  2004b, \mn@doi [\prl] {10.1103/PhysRevLett.93.171104}, 93, 171104

\bibitem[\protect\citeauthoryear{Khrykin et~al.,}{Khrykin et~al.}{2024}]{khrykin_flimflam_2024}
Khrykin I.~S.,  et~al., 2024, \mn@doi [\apj] {10.3847/1538-4357/ad6567}, 973, 151

\bibitem[\protect\citeauthoryear{Kilbinger}{Kilbinger}{2015}]{kilbinger_cosmology_2015}
Kilbinger M.,  2015, \mn@doi [Rep. Prog. Phys.] {10.1088/0034-4885/78/8/086901}, 78, 086901

\bibitem[\protect\citeauthoryear{Kilbinger et~al.,}{Kilbinger et~al.}{2017}]{kilbinger_precision_2017}
Kilbinger M.,  et~al., 2017, \mn@doi [\mnras] {10.1093/mnras/stx2082}, 472, 2126

\bibitem[\protect\citeauthoryear{Kobayashi}{Kobayashi}{2019}]{kobayashi_horndeski_2019}
Kobayashi T.,  2019, \mn@doi [Rep. Prog. Phys.] {10.1088/1361-6633/ab2429}, 82, 086901

\bibitem[\protect\citeauthoryear{Kovacs, {Mao, Sui Ann}, {Basu, Aritra}, {Ma, Yik Ki}, {Pakmor, Ruediger}, {Spitler, Laura G.}  \& {Walker, Charles R. H.}}{Kovacs et~al.}{2024}]{kovacs_2024_dispersion}
Kovacs T.~O.,  {Mao, Sui Ann} {Basu, Aritra} {Ma, Yik Ki} {Pakmor, Ruediger} {Spitler, Laura G.}  {Walker, Charles R. H.} 2024, \mn@doi [A&A] {10.1051/0004-6361/202347459}, 690, A47

\bibitem[\protect\citeauthoryear{{Kumar}, {Shannon}, {Bhandari}  \& {Bannister}}{{Kumar} et~al.}{2021a}]{shannon_craft_20210119}
{Kumar} P.,  {Shannon} R.~M.,  {Bhandari} S.,   {Bannister} K.~W.,  2021a, TNS FRBs, \href {https://ui.adsabs.harvard.edu/abs/2021TNSFR.208....1K} {208, 1}

\bibitem[\protect\citeauthoryear{{Kumar}, {Day}, {Shannon}, {Bhandari}  \& {Qiu}}{{Kumar} et~al.}{2021b}]{kumar_2021_craft}
{Kumar} P.,  {Day} C.~K.,  {Shannon} R.~M.,  {Bhandari} S.,   {Qiu} H.,  2021b, TNS FRBs, \href {https://ui.adsabs.harvard.edu/abs/2021TNSFR2275....1K} {2275, 1}

\bibitem[\protect\citeauthoryear{{LIGO Scientific Collaboration} et~al.,}{{LIGO Scientific Collaboration} et~al.}{2015}]{ligo_2015_advanced}
{LIGO Scientific Collaboration} et~al., 2015, \mn@doi [Classical Quant. Grav.] {10.1088/0264-9381/32/7/074001}, 32, 074001

\bibitem[\protect\citeauthoryear{Lange}{Lange}{2023}]{lange_nautilus_2023}
Lange J.~U.,  2023, \mn@doi [\mnras] {10.1093/mnras/stad2441}, 525, 3181

\bibitem[\protect\citeauthoryear{{Laureijs} et~al.,}{{Laureijs} et~al.}{2011}]{laureijs_euclid_2011}
{Laureijs} R.,  et~al., 2011, {Euclid Definition Study Report}, \mn@doi{10.48550/arXiv.1110.3193}

\bibitem[\protect\citeauthoryear{{Law}}{{Law}}{2023}]{law_dsa_20230810}
{Law} C.~J.,  2023, TNS FRBs, \href {https://ui.adsabs.harvard.edu/abs/2023TNSFR1923....1L} {1923, 1}

\bibitem[\protect\citeauthoryear{{Law}}{{Law}}{2024a}]{law_dsa_20240402}
{Law} C.~J.,  2024a, TNS FRBs, \href {https://ui.adsabs.harvard.edu/abs/2024TNSFR.905....1L} {905, 1}

\bibitem[\protect\citeauthoryear{{Law}}{{Law}}{2024b}]{law_dsa_20240601}
{Law} C.~J.,  2024b, TNS FRBs, \href {https://ui.adsabs.harvard.edu/abs/2024TNSFR1763....1L} {1763, 1}

\bibitem[\protect\citeauthoryear{Leistedt, Peiris, Mortlock, Benoit-Lévy  \& Pontzen}{Leistedt et~al.}{2013}]{leistedt_estimating_2013}
Leistedt B.,  Peiris H.~V.,  Mortlock D.~J.,  Benoit-Lévy A.,   Pontzen A.,  2013, \mn@doi [\mnras] {10.1093/mnras/stt1359}, 435, 1857

\bibitem[\protect\citeauthoryear{Leung \& Collaboration}{Leung \& Collaboration}{2022}]{leung_2022_oneoff}
Leung C.,  Collaboration C.,  2022, Bulletin of the AAS, 54

\bibitem[\protect\citeauthoryear{Li et~al.,}{Li et~al.}{2025}]{li_activerepeating_2025}
Li Y.,  et~al., 2025, {An active repeating fast radio burst in a magnetized eruption environment}, \mn@doi{doi.org/10.48550/arXiv.2503.04727}

\bibitem[\protect\citeauthoryear{Limber}{Limber}{1954}]{limber_analysis_1954}
Limber D.~N.,  1954, \apj, 119, 655

\bibitem[\protect\citeauthoryear{Linder}{Linder}{2017}]{linder_2017_challenges}
Linder E.~V.,  2017, \mn@doi [Phys. Rev. D] {10.1103/PhysRevD.95.023518}, 95, 023518

\bibitem[\protect\citeauthoryear{Lorimer, Bailes, McLaughlin, Narkevic  \& Crawford}{Lorimer et~al.}{2007}]{lorimer_bright_2007}
Lorimer D.~R.,  Bailes M.,  McLaughlin M.~A.,  Narkevic D.~J.,   Crawford F.,  2007, \mn@doi [Science] {10.1126/science.1147532}, 318, 777

\bibitem[\protect\citeauthoryear{Ma \& Bertschinger}{Ma \& Bertschinger}{1995}]{ma_cosmological_1995}
Ma C.-P.,  Bertschinger E.,  1995, \mn@doi [Astrophys. J.] {10.1086/176550}, 455, 7

\bibitem[\protect\citeauthoryear{Macquart et~al.,}{Macquart et~al.}{2020}]{macquart_census_2020}
Macquart J.-P.,  et~al., 2020, \mn@doi [Nature] {10.1038/s41586-020-2300-2}, 581, 391

\bibitem[\protect\citeauthoryear{Masui \& Sigurdson}{Masui \& Sigurdson}{2015}]{masui_dispersion_2015}
Masui K.~W.,  Sigurdson K.,  2015, \mn@doi [Phys. Rev. Lett.] {10.1103/PhysRevLett.115.121301}, 115, 121301

\bibitem[\protect\citeauthoryear{Mead, Peacock, Heymans, Joudaki  \& Heavens}{Mead et~al.}{2015}]{mead_accurate_2015}
Mead A.~J.,  Peacock J.~A.,  Heymans C.,  Joudaki S.,   Heavens A.~F.,  2015, \mn@doi [\mnras] {10.1093/mnras/stv2036}, 454, 1958

\bibitem[\protect\citeauthoryear{Mead, Tröster, Heymans, Van~Waerbeke  \& McCarthy}{Mead et~al.}{2020}]{mead_hydrodynamical_2020}
Mead A.~J.,  Tröster T.,  Heymans C.,  Van~Waerbeke L.,   McCarthy I.~G.,  2020, \mn@doi [Astron. Astrophys.] {10.1051/0004-6361/202038308}, 641, A130

\bibitem[\protect\citeauthoryear{Mo, Zhu, Wang, Tang  \& Feng}{Mo et~al.}{2022}]{Mo_2022_dispersion}
Mo J.-F.,  Zhu W.,  Wang Y.,  Tang L.,   Feng L.-L.,  2022, \mn@doi [\mnras] {10.1093/mnras/stac3104}, 518, 539

\bibitem[\protect\citeauthoryear{Mu\~noz \& Loeb}{Mu\~noz \& Loeb}{2018}]{mu_2018_finding}
Mu\~noz J.~B.,  Loeb A.,  2018, \mn@doi [Phys. Rev. D] {10.1103/PhysRevD.98.103518}, 98, 103518

\bibitem[\protect\citeauthoryear{Nicola, García-García, Alonso, Dunkley, Ferreira, Slosar  \& Spergel}{Nicola et~al.}{2021}]{nicola_cosmic_2021}
Nicola A.,  García-García C.,  Alonso D.,  Dunkley J.,  Ferreira P.~G.,  Slosar A.,   Spergel D.~N.,  2021, \mn@doi [\jcap] {10.1088/1475-7516/2021/03/067}, 2021, 067

\bibitem[\protect\citeauthoryear{{Niu} et~al.,}{{Niu} et~al.}{2021}]{niu_transient_20210925}
{Niu} C.,  et~al., 2021, TNS FRBs, \href {https://ui.adsabs.harvard.edu/abs/2021TNSFR3302....1N} {3302, 1}

\bibitem[\protect\citeauthoryear{Noller}{Noller}{2020}]{noller_2020_cosmological}
Noller J.,  2020, \mn@doi [Phys. Rev. D] {10.1103/PhysRevD.101.063524}, 101, 063524

\bibitem[\protect\citeauthoryear{{Petroff}}{{Petroff}}{2020}]{petroff_frbcat_20200812}
{Petroff} E.,  2020, TNS FRBs, \href {https://ui.adsabs.harvard.edu/abs/2020TNSFR2470....1P} {2470, 1}

\bibitem[\protect\citeauthoryear{Petroff, Hessels  \& Lorimer}{Petroff et~al.}{2019}]{petroff_fast_2019}
Petroff E.,  Hessels J. W.~T.,   Lorimer D.~R.,  2019, \mn@doi [Astron. Astrophys. Rev.] {10.1007/s00159-019-0116-6}, 27, 4

\bibitem[\protect\citeauthoryear{{Planck Collaboration} et~al.,}{{Planck Collaboration} et~al.}{2016}]{planck_collaboration_xiv_planck_2015}
{Planck Collaboration} et~al., 2016, \mn@doi [A&A] {10.1051/0004-6361/201525814}, 594, A14

\bibitem[\protect\citeauthoryear{{Planck Collaboration} et~al.,}{{Planck Collaboration} et~al.}{2020a}]{planck_2020_overview}
{Planck Collaboration} et~al., 2020a, \mn@doi [A&A] {10.1051/0004-6361/201833880}, 641, A1

\bibitem[\protect\citeauthoryear{{Planck Collaboration} et~al.,}{{Planck Collaboration} et~al.}{2020b}]{planck_collaboration_planck_2020}
{Planck Collaboration} et~al., 2020b, \mn@doi [A\&A] {10.1051/0004-6361/201833910}, 641, A6

\bibitem[\protect\citeauthoryear{Prochaska et~al.,}{Prochaska et~al.}{2019}]{prochaska_low_2019}
Prochaska J.~X.,  et~al., 2019, \mn@doi [Science] {10.1126/science.aay0073}, 366, 231

\bibitem[\protect\citeauthoryear{Rafiei-Ravandi et~al.,}{Rafiei-Ravandi et~al.}{2021}]{rafiei-ravandi_chimefrb_2021}
Rafiei-Ravandi M.,  et~al., 2021, \mn@doi [\apj] {10.3847/1538-4357/ac1dab}, 922, 42

\bibitem[\protect\citeauthoryear{Ratra}{Ratra}{1988}]{ratra_expressions_1988}
Ratra B.,  1988, \mn@doi [\prd] {10.1103/PhysRevD.38.2399}, 38, 2399

\bibitem[\protect\citeauthoryear{Ratra \& Peebles}{Ratra \& Peebles}{1988}]{ratra_1988_quintessence}
Ratra B.,  Peebles P. J.~E.,  1988, \mn@doi [Phys. Rev. D] {10.1103/PhysRevD.37.3406}, 37, 3406

\bibitem[\protect\citeauthoryear{{Ravi}}{{Ravi}}{2023}]{ravi_dsa_20230816}
{Ravi} V.,  2023, TNS FRBs, \href {https://ui.adsabs.harvard.edu/abs/2023TNSFR1990....1R} {1990, 1}

\bibitem[\protect\citeauthoryear{Ravi et~al.,}{Ravi et~al.}{2019}]{ravi_fast_2019}
Ravi V.,  et~al., 2019, \mn@doi [Nature] {10.1038/s41586-019-1389-7}, 572, 352

\bibitem[\protect\citeauthoryear{Reischke \& Hagstotz}{Reischke \& Hagstotz}{2023a}]{2023_reischke_consistent}
Reischke R.,  Hagstotz S.,  2023a, \mn@doi [MNRAS] {10.1093/mnras/stad1866}, 523, 6264

\bibitem[\protect\citeauthoryear{Reischke \& Hagstotz}{Reischke \& Hagstotz}{2023b}]{reischke_cosmological_2023}
Reischke R.,  Hagstotz S.,  2023b, \mn@doi [\mnras] {10.1093/mnras/stad1645}, 524, 2237

\bibitem[\protect\citeauthoryear{Reischke, Mancini, Schäfer  \& Merkel}{Reischke et~al.}{2019}]{reischke_investigating_2019}
Reischke R.,  Mancini A.~S.,  Schäfer B.~M.,   Merkel P.~M.,  2019, \mn@doi [\mnras] {10.1093/mnras/sty2919}, 482, 3274

\bibitem[\protect\citeauthoryear{Reischke, Hagstotz  \& Lilow}{Reischke et~al.}{2021}]{reischke_probing_2021}
Reischke R.,  Hagstotz S.,   Lilow R.,  2021, \mn@doi [Phys. Rev. D] {10.1103/PhysRevD.103.023517}, 103, 023517

\bibitem[\protect\citeauthoryear{Reischke, Hagstotz  \& Lilow}{Reischke et~al.}{2022}]{reischke_consistent_2022}
Reischke R.,  Hagstotz S.,   Lilow R.,  2022, \mn@doi [\mnras] {10.1093/mnras/stab3571}, 512, 285

\bibitem[\protect\citeauthoryear{Reischke, Neumann, Bertmann, Hagstotz  \& Hildebrandt}{Reischke et~al.}{2023}]{reischke_calibrating_2023}
Reischke R.,  Neumann D.,  Bertmann K.~A.,  Hagstotz S.,   Hildebrandt H.,  2023, Calibrating baryonic feedback with weak lensing and fast radio bursts, \mn@doi{10.48550/arXiv.2309.09766}

\bibitem[\protect\citeauthoryear{Reischke, Kovač, Nicola, Hagstotz  \& Schneider}{Reischke et~al.}{2024}]{reischke_2024_analytical}
Reischke R.,  Kovač M.,  Nicola A.,  Hagstotz S.,   Schneider A.,  2024, An analytical model for the dispersion measure of Fast Radio Burst host galaxies, \mn@doi{10.48550/arXiv.2411.17682}

\bibitem[\protect\citeauthoryear{Riess, Filippenko, Challis  \& {et al.}}{Riess et~al.}{1998}]{riess_observational_1998}
Riess A.~G.,  Filippenko A.~V.,  Challis P.,   {et al.} 1998, \mn@doi [\aj] {10.1086/300499}, 116, 1009

\bibitem[\protect\citeauthoryear{Sawicki \& Bellini}{Sawicki \& Bellini}{2015}]{sawicki_2015_limits}
Sawicki I.,  Bellini E.,  2015, \mn@doi [Phys. Rev. D] {10.1103/PhysRevD.92.084061}, 92, 084061

\bibitem[\protect\citeauthoryear{Secco et~al.,}{Secco et~al.}{2022}]{secco_2022_DESY3}
Secco L.~F.,  et~al., 2022, \mn@doi [Phys. Rev. D] {10.1103/PhysRevD.105.023515}, 105, 023515

\bibitem[\protect\citeauthoryear{{Shannon}}{{Shannon}}{2022a}]{shannon_craft_20220612}
{Shannon} R.~M.,  2022a, TNS FRBs, \href {https://ui.adsabs.harvard.edu/abs/2022TNSFR1658....1S} {1658, 1}

\bibitem[\protect\citeauthoryear{{Shannon}}{{Shannon}}{2022b}]{shannon_craft_20220726}
{Shannon} R.~M.,  2022b, TNS FRBs, \href {https://ui.adsabs.harvard.edu/abs/2022TNSFR2108....1S} {2108, 1}

\bibitem[\protect\citeauthoryear{{Shannon}}{{Shannon}}{2022c}]{shannon_craft_20220919}
{Shannon} R.~M.,  2022c, TNS FRBs, \href {https://ui.adsabs.harvard.edu/abs/2022TNSFR2708....1S} {2708, 1}

\bibitem[\protect\citeauthoryear{{Shannon}}{{Shannon}}{2022d}]{shannon_craft_20221113}
{Shannon} R.~M.,  2022d, TNS FRBs, \href {https://ui.adsabs.harvard.edu/abs/2022TNSFR3302....1S} {3302, 1}

\bibitem[\protect\citeauthoryear{{Shannon}}{{Shannon}}{2023}]{shannon_craft_20230204}
{Shannon} R.~M.,  2023, TNS FRBs, \href {https://ui.adsabs.harvard.edu/abs/2023TNSFR.287....1S} {287, 1}

\bibitem[\protect\citeauthoryear{{Shannon} \& {Kumar}}{{Shannon} \& {Kumar}}{2022}]{shannon_craft_20220126}
{Shannon} R.~M.,  {Kumar} P.,  2022, TNS FRBs, \href {https://ui.adsabs.harvard.edu/abs/2022TNSFR.215....1S} {215, 1}

\bibitem[\protect\citeauthoryear{{Shannon} \& {Uttarkar}}{{Shannon} \& {Uttarkar}}{2023a}]{shannon_20230527}
{Shannon} R.~M.,  {Uttarkar} P.~A.,  2023a, TNS FRBs, \href {https://ui.adsabs.harvard.edu/abs/2023TNSFR1251....1S} {1251, 1}

\bibitem[\protect\citeauthoryear{{Shannon} \& {Uttarkar}}{{Shannon} \& {Uttarkar}}{2023b}]{Shannon_202307}
{Shannon} R.~M.,  {Uttarkar} P.~A.,  2023b, TNS FRBs, \href {https://ui.adsabs.harvard.edu/abs/2023TNSFR1609....1S} {1609, 1}

\bibitem[\protect\citeauthoryear{{Shannon} \& {Uttarkar}}{{Shannon} \& {Uttarkar}}{2023c}]{shannon_craft_20230902}
{Shannon} R.~M.,  {Uttarkar} P.~A.,  2023c, TNS FRBs, \href {https://ui.adsabs.harvard.edu/abs/2023TNSFR2149....1S} {2149, 1}

\bibitem[\protect\citeauthoryear{{Shannon} \& {Uttarkar}}{{Shannon} \& {Uttarkar}}{2024a}]{Shannon_craft_20240112}
{Shannon} R.~M.,  {Uttarkar} P.~A.,  2024a, TNS FRBs, \href {https://ui.adsabs.harvard.edu/abs/2024TNSFR.130....1S} {130, 1}

\bibitem[\protect\citeauthoryear{{Shannon} \& {Uttarkar}}{{Shannon} \& {Uttarkar}}{2024b}]{Shannon_20240712}
{Shannon} R.~M.,  {Uttarkar} P.~A.,  2024b, TNS FRBs, \href {https://ui.adsabs.harvard.edu/abs/2024TNSFR2411....1S} {2411, 1}

\bibitem[\protect\citeauthoryear{{Shannon}, {Kumar}, {Bhandari}  \& {Macquart}}{{Shannon} et~al.}{2019}]{shannon_2019_ASKAP191001}
{Shannon} R.~M.,  {Kumar} P.,  {Bhandari} S.,   {Macquart} J.~P.,  2019, The Astronomer's Telegram, \href {https://ui.adsabs.harvard.edu/abs/2019ATel13166....1S} {13166, 1}

\bibitem[\protect\citeauthoryear{Sherman et~al.,}{Sherman et~al.}{2024}]{sherman_2024_deep}
Sherman M.~B.,  et~al., 2024, \mn@doi [\apj] {10.3847/1538-4357/ad275e}, 964, 131

\bibitem[\protect\citeauthoryear{Shirasaki, Kashiyama  \& Yoshida}{Shirasaki et~al.}{2017}]{shirasaki_large-scale_2017}
Shirasaki M.,  Kashiyama K.,   Yoshida N.,  2017, \mn@doi [\prd] {10.1103/PhysRevD.95.083012}, 95, 083012

\bibitem[\protect\citeauthoryear{Shull, Smith  \& Danforth}{Shull et~al.}{2012}]{shull_baryon_2012}
Shull J.~M.,  Smith B.~D.,   Danforth C.~W.,  2012, \mn@doi [\apj] {10.1088/0004-637X/759/1/23}, 759, 23

\bibitem[\protect\citeauthoryear{Spurio~Mancini, Taylor, Reischke, Kitching, Pettorino, Schäfer, Zieser  \& Merkel}{Spurio~Mancini et~al.}{2018a}]{spurio_mancini_3d_2018}
Spurio~Mancini A.,  Taylor P.~L.,  Reischke R.,  Kitching T.,  Pettorino V.,  Schäfer B.~M.,  Zieser B.,   Merkel P.~M.,  2018a, \mn@doi [Phys. Rev. D] {10.1103/PhysRevD.98.103507}, 98, 103507

\bibitem[\protect\citeauthoryear{Spurio~Mancini, Reischke, Pettorino, Schäfer  \& Zumalacárregui}{Spurio~Mancini et~al.}{2018b}]{spurio_mancini_testing_2018}
Spurio~Mancini A.,  Reischke R.,  Pettorino V.,  Schäfer B.~M.,   Zumalacárregui M.,  2018b, \mn@doi [\mnras] {10.1093/mnras/sty2092}, 480, 3725

\bibitem[\protect\citeauthoryear{Spurio~Mancini et~al.,}{Spurio~Mancini et~al.}{2019}]{spurio_mancini_kids_2019}
Spurio~Mancini A.,  et~al., 2019, \mn@doi [\mnras] {10.1093/mnras/stz2581}, 490, 2155

\bibitem[\protect\citeauthoryear{Spurio~Mancini, Piras, Alsing, Joachimi  \& Hobson}{Spurio~Mancini et~al.}{2022}]{spuriomancini_cosmopower_2022}
Spurio~Mancini A.,  Piras D.,  Alsing J.,  Joachimi B.,   Hobson M.~P.,  2022, \mn@doi [\mnras] {10.1093/mnras/stac064}, 511, 1771

\bibitem[\protect\citeauthoryear{Takahashi, Ioka, Mori  \& Funahashi}{Takahashi et~al.}{2021}]{takahashi_statistical_2021}
Takahashi R.,  Ioka K.,  Mori A.,   Funahashi K.,  2021, \mn@doi [\mnras] {10.1093/mnras/stab170}, 502, 2615

\bibitem[\protect\citeauthoryear{Tang, Lin  \& Li}{Tang et~al.}{2023}]{Tang_2023_inferring}
Tang L.,  Lin H.-N.,   Li X.,  2023, \mn@doi [Chinese Physics C] {10.1088/1674-1137/acda1c}, 47, 085105

\bibitem[\protect\citeauthoryear{Tanidis \& Camera}{Tanidis \& Camera}{2019}]{tanidis_2019_developing}
Tanidis K.,  Camera S.,  2019, \mn@doi [\mnras] {10.1093/mnras/stz2366}, 489, 3385

\bibitem[\protect\citeauthoryear{Tegmark, Taylor  \& Heavens}{Tegmark et~al.}{1997}]{tegmark_karhunen-loeve_1997}
Tegmark M.,  Taylor A.~N.,   Heavens A.~F.,  1997, \mn@doi [\apj] {10.1086/303939}, 480, 22

\bibitem[\protect\citeauthoryear{Tiesinga, Mohr, Newell  \& Taylor}{Tiesinga et~al.}{2021}]{tiesinga_codata_2021}
Tiesinga E.,  Mohr P.~J.,  Newell D.~B.,   Taylor B.~N.,  2021, \mn@doi [Rev. Mod. Phys.] {10.1103/RevModPhys.93.025010}, 93, 025010

\bibitem[\protect\citeauthoryear{Tripathi, Sangwan  \& Jassal}{Tripathi et~al.}{2017}]{Tripathi_2017_dark}
Tripathi A.,  Sangwan A.,   Jassal H.,  2017, \mn@doi [\jcap] {10.1088/1475-7516/2017/06/012}, 2017, 012

\bibitem[\protect\citeauthoryear{Tröster et~al.,}{Tröster et~al.}{2022}]{tröster_2022_joint}
Tröster T.,  et~al., 2022, \mn@doi [A&A] {10.1051/0004-6361/202142197}, 660, A27

\bibitem[\protect\citeauthoryear{Vainshtein}{Vainshtein}{1972}]{vainshtein_problem_1972}
Vainshtein A.~I.,  1972, \mn@doi [Physics Letters B] {10.1016/0370-2693(72)90147-5}, 39, 393

\bibitem[\protect\citeauthoryear{Vanderlinde et~al.,}{Vanderlinde et~al.}{2019}]{vanderlinde_2019_chord}
Vanderlinde K.,  et~al., 2019, The Canadian Hydrogen Observatory and Radio- transient Detector (CHORD), \mn@doi{10.5281/zenodo.3765414}

\bibitem[\protect\citeauthoryear{Walters, Weltman, Gaensler, Ma  \& Witzemann}{Walters et~al.}{2018}]{Walters_2018_future}
Walters A.,  Weltman A.,  Gaensler B.~M.,  Ma Y.-Z.,   Witzemann A.,  2018, \mn@doi [\apj] {10.3847/1538-4357/aaaf6b}, 856, 65

\bibitem[\protect\citeauthoryear{Weinberg, Mortonson, Eisenstein, Hirata, Riess  \& Rozo}{Weinberg et~al.}{2013}]{weinberg_observational_2013}
Weinberg D.~H.,  Mortonson M.~J.,  Eisenstein D.~J.,  Hirata C.,  Riess A.~G.,   Rozo E.,  2013, \mn@doi [\physrep] {10.1016/j.physrep.2013.05.001}, 530, 87

\bibitem[\protect\citeauthoryear{Wojtak, Hansen  \& Hjorth}{Wojtak et~al.}{2011}]{wojtak_gravitational_2011}
Wojtak R.,  Hansen S.~H.,   Hjorth J.,  2011, \mn@doi [Nature] {10.1038/nature10445}, 477, 567

\bibitem[\protect\citeauthoryear{Wu, Zhang  \& Wang}{Wu et~al.}{2022}]{wu_8_2022}
Wu Q.,  Zhang G.-Q.,   Wang F.-Y.,  2022, \mn@doi [\mnras] {10.1093/mnrasl/slac022}, 515, L1

\bibitem[\protect\citeauthoryear{Xu et~al.,}{Xu et~al.}{2023}]{xu_2023_blinkverse}
Xu J.,  et~al., 2023, \mn@doi [Universe] {10.3390/universe9070330}, 9

\bibitem[\protect\citeauthoryear{Yamasaki \& Totani}{Yamasaki \& Totani}{2020}]{yamasaki_galactic_2020}
Yamasaki S.,  Totani T.,  2020, \mn@doi [ApJ] {10.3847/1538-4357/ab58c4}, 888, 105

\bibitem[\protect\citeauthoryear{Yang \& Zhang}{Yang \& Zhang}{2016}]{yang_2016_extracting}
Yang Y.-P.,  Zhang B.,  2016, \mn@doi [\apjl] {10.3847/2041-8205/830/2/L31}, 830, L31

\bibitem[\protect\citeauthoryear{Yao, Manchester  \& Wang}{Yao et~al.}{2017}]{yao_new_2017}
Yao J.~M.,  Manchester R.~N.,   Wang N.,  2017, \mn@doi [\apj] {10.3847/1538-4357/835/1/29}, 835, 29

\bibitem[\protect\citeauthoryear{Zhang, Yu, He  \& Wang}{Zhang et~al.}{2020}]{Zhang_2020_dispersion}
Zhang G.~Q.,  Yu H.,  He J.~H.,   Wang F.~Y.,  2020, \mn@doi [\apj] {10.3847/1538-4357/abaa4a}, 900, 170

\bibitem[\protect\citeauthoryear{Zhang et~al.,}{Zhang et~al.}{2024}]{zhang_2024_BINGO}
Zhang X.,  et~al., 2024, The BINGO/ABDUS Project: Forecast for cosmological parameter from a mock Fast Radio Bursts survey, \mn@doi{10.48550/arXiv.2411.17516}

\bibitem[\protect\citeauthoryear{Zhou, Li, Wang, Fan  \& Wei}{Zhou et~al.}{2014}]{zhou_fast_2014}
Zhou B.,  Li X.,  Wang T.,  Fan Y.-Z.,   Wei D.-M.,  2014, \mn@doi [Phys. Rev. D] {10.1103/PhysRevD.89.107303}, 89, 107303

\bibitem[\protect\citeauthoryear{Zumalacárregui, Bellini, Sawicki, Lesgourgues  \& Ferreira}{Zumalacárregui et~al.}{2017}]{zumalacarregui_hiclass:_2016}
Zumalacárregui M.,  Bellini E.,  Sawicki I.,  Lesgourgues J.,   Ferreira P.~G.,  2017, \mn@doi [\jcap] {10.1088/1475-7516/2017/08/019}, 2017, 019

\bibitem[\protect\citeauthoryear{{Zwaniga} \& {Kaspi}}{{Zwaniga} \& {Kaspi}}{2021}]{zwaniga_chimefrb_discovery_2021}
{Zwaniga} A.~V.,  {Kaspi} V.~M.,  2021, TNS FRBs, \href {https://ui.adsabs.harvard.edu/abs/2021TNSFR2007....1Z} {2007, 1}

\makeatother
\end{thebibliography}

\renewcommand\thesection{\alpha{section}}
\renewcommand\thesubsection{\thesection.\arabic{subsection}}
\renewcommand\thefigure{\thesection.\arabic{figure}} 
\setcounter{figure}{0}

\cleardoublepage

\appendix

\section{Additional constraints from Planck CMB}
\label{ch:app_adding_planck}

{We additionally investigate the influence of adding information from the \textsc{Planck2018}\,\citep{planck_2020_overview} temperature anisotropy power spectrum. Based on the chosen $\propto \Omega_\Lambda$ parametrisation used for modified gravity, we expect the main constraining power thereof to originate from degeneracy breaking between cosmological parameters. The posteriors are obtained by sampling the combined \textsc{Planck2018}-$\mathrm{TT}+${Euclid-CS}$\times$FRB likelihood, whereas the Planck covariance is obtained from inverting the corresponding Fisher matrix and marginalising over nuisance parameters not included in the {Euclid-CS}$\times$FRB likelihood.}

\begin{figure}[ht]
    \centering
    \includegraphics[width=1\linewidth]{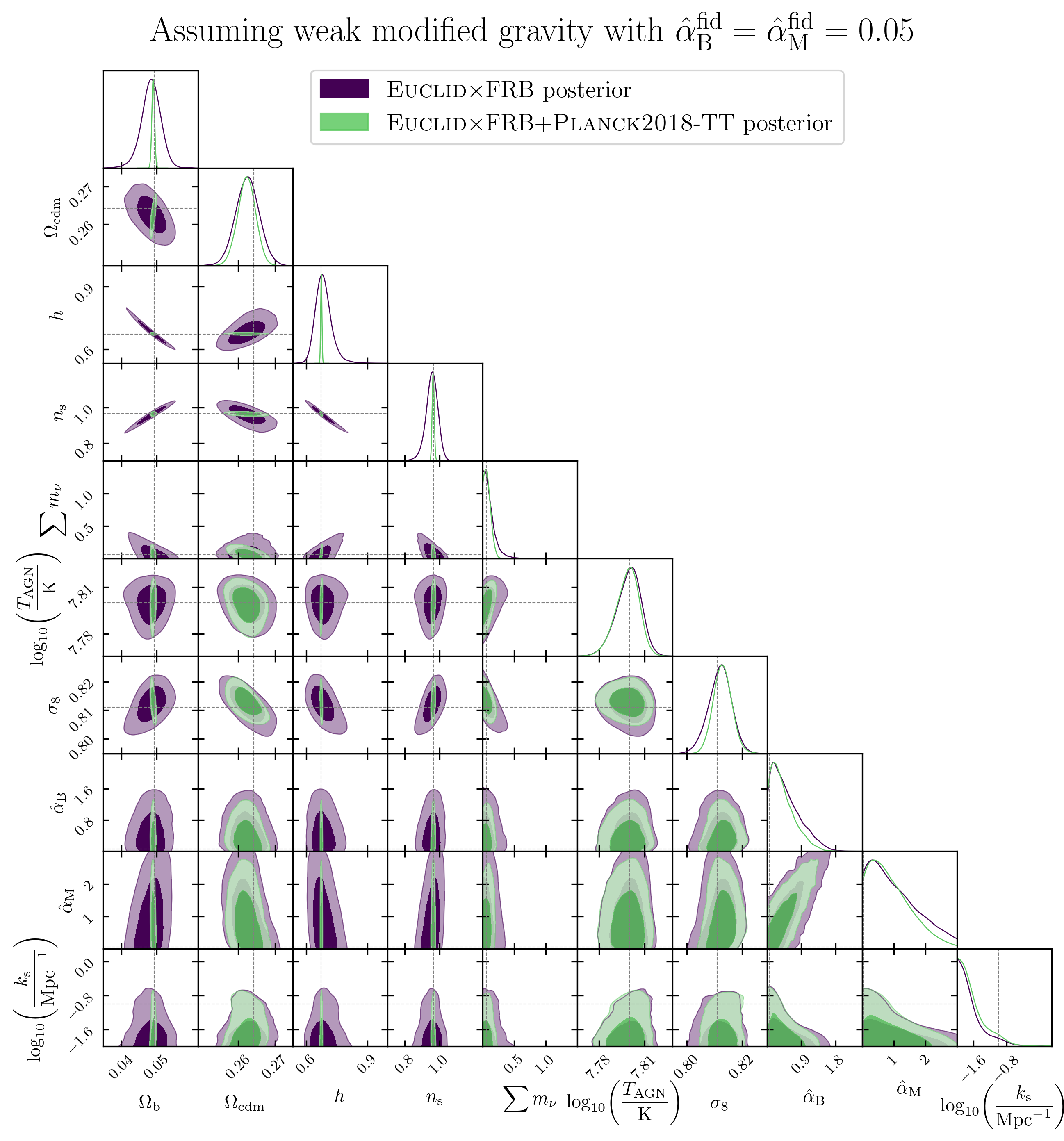}
    \caption{68\% and 95 \% posterior confidence regions (dark and light areas, respectively) for \textsc{Euclid-CS}$\times$FRB cross-correlation alone and in combination with \textsc{Planck2018} temperature anisotropy auto-correlation. The fiducial cosmology taken from Tab. \ref{tab:data_fid_cosmology} is indicated by grey, dashed lines.}
    \label{fig:app_euclidxfrb_plus_plancktt}
\end{figure}

Furthermore, the Planck likelihood carries no information about the Horndeski screening scale and the baryonic feedback strength, hence we extend the Fisher matrix for $\Tagn$ and $\ks$ with small values ($10^{-5}$) along the corresponding diagonal, thus implying no intrinsic constraining power on these parameters. The posterior contours for the weak modified gravity case are shown in Fig. \ref{fig:app_euclidxfrb_plus_plancktt}. While there is a substantial increase in contraining power on $\Omega_\mathrm{b}$, the 68\,\% confidence region for $h$, and $n_s$, the $\aB$ ($\aM$) decreases only by a factor of \textsb{1.23 (1.15)}. The effect is even less for the strong modified gravity case, where the error margin decreases by a factor of \textsb{1.01} (1.04) for $\aB$ ($\aM$). Therefore, adding CMB temperature fluctuation information has only diminishing returns regarding \textsb{this modified gravity parametrisation}.

\section{Accuracy and details of the \texttt{cosmopower} emulator}
\label{ch:app_accuracy_emulator}
For the emulator training, we use five learning steps with subsequent learning rates of $10^{-2}$, $10^{-3}$, $10^{-4}$, $10^{-5}$, $10^{-6}$. For each, the number of epochs range from a minimum of 100 for early stopping due to lack of model improvement to a maximum of 1000. Ten per cent of the data is used for validation. The batch sizes and hidden layer neuron structure for all models are depicted in Tab. \ref{tab:data_nn_params}.
 \renewcommand{\arraystretch}{1.15}
 \begin{table}[ht]
	\centering
 	\begin{tabular}{lcc}
	\hline
	\multicolumn{1}{c}{model} & batch sizes & hidden neuron structure \\
	\hline
	$P_\mathrm{mm}(k)$ & 256 & 200$\times$200 \\
	$b_\mathrm{e}(k)$ & 256 & 200$\times$200 \\
	$\eta(k)$ & 512 & 200$\times$200 \\
	$\mu(k)$ & 512 & 200$\times$200 \\
	$\chi(z)$ & 512 & 30$\times$70 \\
	\hline
	\end{tabular}
	\caption{\texttt{cosmopower} Emulator training settings for the $\{P_\mathrm{mm}(k),\,b_\mathrm{e}(k),\,\eta(k),\,\mu(k),\,\chi(z)\}$ models.}
 \label{tab:data_nn_params}
\end{table}
An analysis of the test data reveals the emulators to be accurate within 5\% (see Fig.\;\ref{fig:app_cosmopower_accuracy} for more information). 

\begin{figure}[ht!]
    \centering
    \includegraphics[width=.7\linewidth]{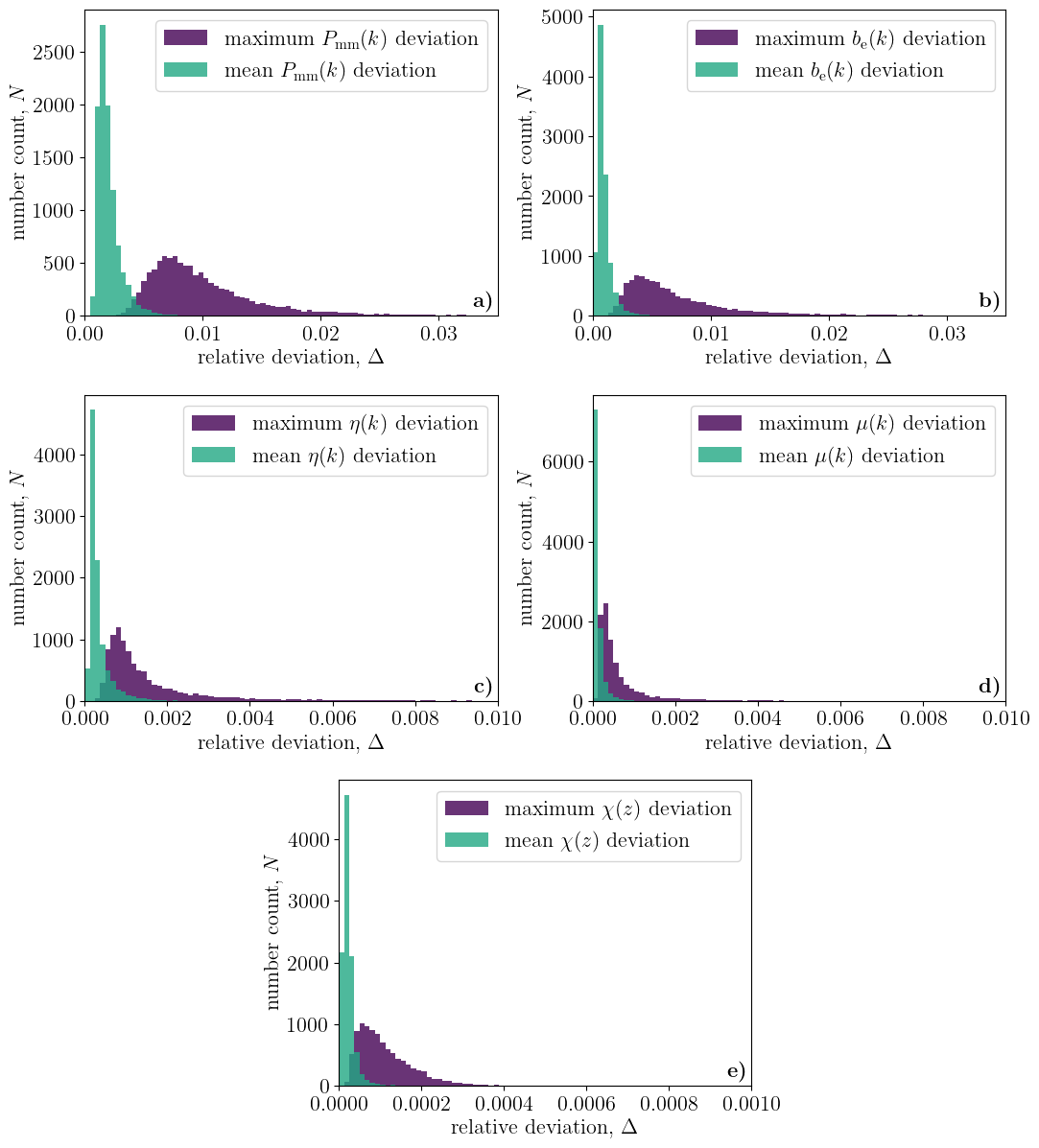}
    \caption{Histogram of the maximum and mean deviation of the emulated \textbf{a)} $P_\mathrm{mm}(k)$, \textbf{b)} $b_\mathrm{e}(k)$, \textbf{c)} $\eta(k)$, \textbf{d)} $\mu(k)$, \textbf{e)} $\chi(z)$ to the simulated test data.} 
    \label{fig:app_cosmopower_accuracy}
    \vspace{.25cm}
\end{figure}

We use 9793 samples to test the accuracy of the emulator. To evaluate the trained neural networks, the five functions $f_i\in\{P_\mathrm{mm}(k),\,b_\mathrm{e}(k),\,\eta(k),$ $\mu(k),\,\chi(z)\}$ are emulated with the same cosmological parameters used for calculating the test data set. Then, the relative deviation to the validation data is calculated as
\begin{align}
	\Delta := \left|1-\frac{f_i^{\texttt{cosmopower}}}{f_i^{\texttt{hi\_class}}}\right| 
\end{align}
for each mode, $k$ or $z$, depending on the function. Finally, by taking a histogram of the maximum and mean over all modes for each set of input parameters, a visual representation of how the neural network performs at worst and on average is created, shown in Fig.\;\ref{fig:app_cosmopower_accuracy}. At worst, the emulator deviates $\lesssim 3\%$ and performs much better most of the time. Therefore, we exclude the emulator as a dominating systematic in this work.

\section{MCMC convergence test}
\setcounter{figure}{0}
\label{ch:app_convergence_test}

The convergence tests are conducted by varying the hyperparameters of \texttt{Nautilus}-sampler: the maximum fraction of the evidence contained in the live set before the exploration phase ends, ``f\_live'', and the minimum demanded effective sample size, ``n\_eff''. In Fig\;\ref{fig:app_convergence_test} we show the results for the \textsc{Euclid-CS} auto-correlations MCMCs with the fiducial cosmology given in Tab\;\ref{tab:data_fid_cosmology}. Note that varying the hyperparameters has very little effect on the resulting contours, hence we use the \texttt{Nautilus} standard settings. Similar behaviour is obtained when testing the {Euclid-CS}$\times$FRB cross-correlation posterior for convergence.

\begin{figure}[ht]
    \centering
    \includegraphics[width=1\linewidth]{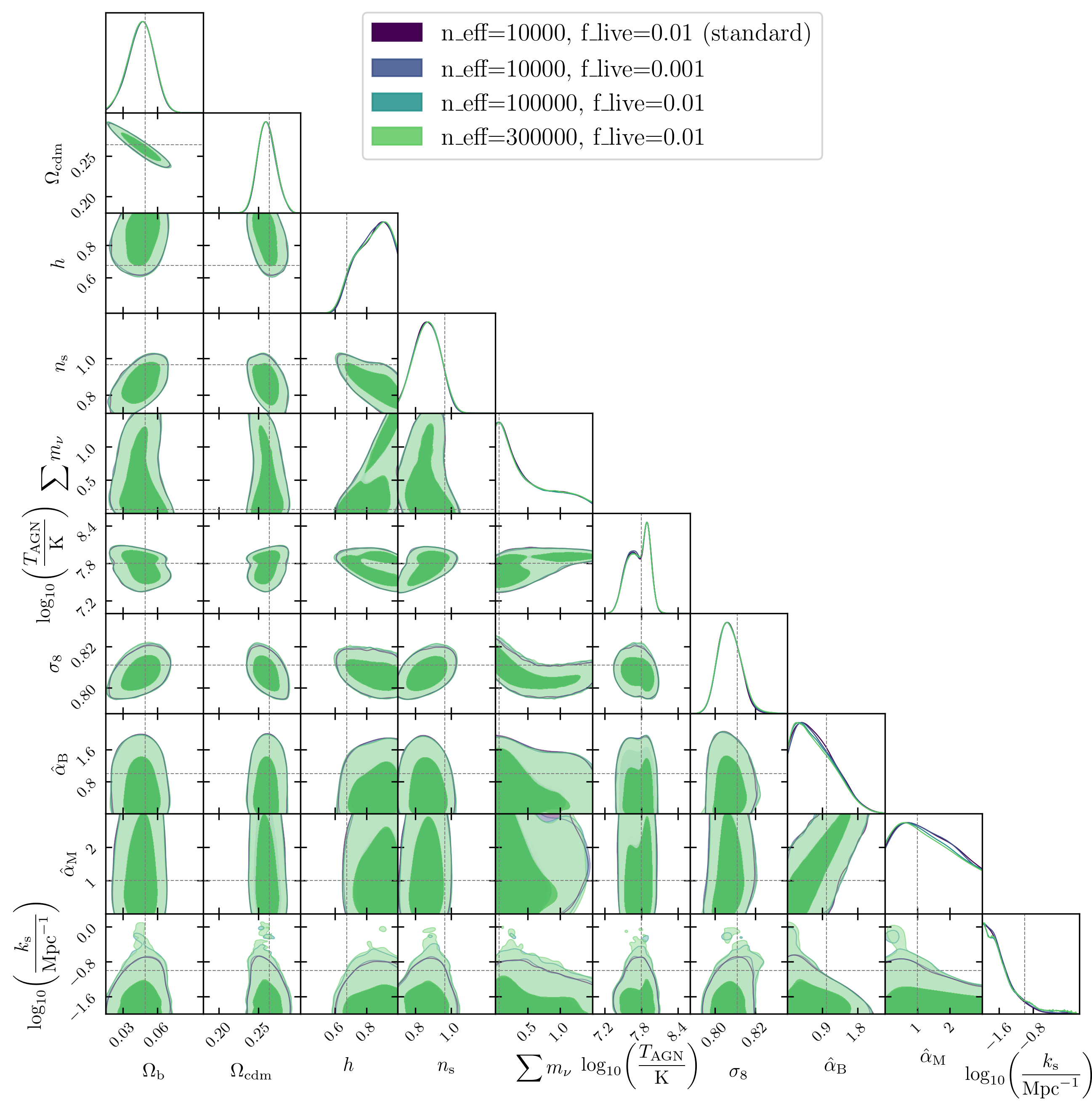}
    \caption{\textsc{Euclid-CS} auto-correlation posterior contours for different settings of \texttt{Nautilus}-sampler.}
    \label{fig:app_convergence_test}
\end{figure}

\clearpage

\section{FRBs used for real life application}

In Tab. \ref{tab:app_frb_list} we summarise all the FRBs used in Section \ref{ch:appli} \textsb{sorted by ascending redshift}:

\renewcommand{\arraystretch}{1.4}
\begin{longtable}{c|cccccc}
\caption{FRBs used for the analysis presented in Section \ref{ch:appli}.}
\label{tab:app_frb_list} \\

\hline
FRB & $\mathrm{DM}_\mathrm{obs}$ & z & $\mathrm{DM}_\mathrm{MW}$ & ra[rad] & dec[rad] & Source \\
\hline
\endfirsthead

\hline
FRB & $\mathrm{DM}_\mathrm{obs}$ & z & $\mathrm{DM}_\mathrm{MW}$ & ra[rad] & dec[rad] & Source \\
\hline
\endhead

\hline
\endfoot

\hline
\endlastfoot

20230708A  &  411.51  &  0.105  &  50.0  &  -2.7889  &  -0.9661 & \citet{Shannon_202307} \\ 
20191106C  &  332.2  &  0.10775  &  25.0  &  -2.9094  &  0.7505 & \citet{chime_2021_first}\\ 
20220914A  &  631.28  &  0.1139  &  55.2  &  -2.8134  &  1.28 & \citet{sherman_2024_deep} \\ 
20190608  &  339.5  &  0.11778  &  37.2  &  -2.7529  &  -0.1378 & \citet{zwaniga_chimefrb_discovery_2021} \\ 
20190110C  &  221.6  &  0.12244  &  37.1  &  -2.8515  &  0.7233 & \citet{zwaniga_chimefrb_discovery_2021} \\ 
20240310A  &  601.8  &  0.127  &  36.0  &  -3.1182  &  -0.7756 & \citet{Shannon_20240712} \\ 
20240213A  &  357.4  &  0.1185  &  32.1  &  -2.9482  &  1.2929 & \citet{law_dsa_20240601} \\ 
20230628A  &  345.15  &  0.1265  &  30.83  &  -2.9475  &  1.2616 & \citet{law_dsa_20240601} \\ 
20210807D  &  251.3  &  0.1293  &  121.2  &  -2.7934  &  -0.0133 & \citet{james_measurement_2022}\\ 
20240114A  &  527.7  &  0.13  &  38.82  &  -2.767  &  0.0756 & \citet{abbott_chimefrb_20240202} \\ 
20210410D  &  578.78  &  0.1415  &  56.2  &  -2.7622  &  -1.3844 & \citet{caleb_2023_subarcsec} \\ 
20231226A  &  329.9  &  0.1569  &  145.0  &  -2.9608  &  0.1066 & \citet{Shannon_craft_20240112} \\ 
20230526A  &  361.4  &  0.157  &  50.0  &  -3.1157  &  -0.9201 & \citet{shannon_20230527} \\ 
20220920A  &  314.99  &  0.158239  &  40.3  &  -2.862  &  1.2378 & \citet{sherman_2024_deep} \\ 
20200430  &  380.25  &  0.16  &  27.0  &  -2.8743  &  0.216 & \citet{kumar_2021_craft} \\ 
20210603A  &  500.147  &  0.177  &  40.0  &  -3.1296  &  0.3705 & \citet{leung_2022_oneoff} \\ 
20220529A  &  246.0  &  0.1839  &  30.92  &  -3.1194  &  0.3601 & \citet{li_activerepeating_2025} \\ 
20220725A  &  290.4  &  0.1926  &  31.0  &  -2.7305  &  -0.6281 & \citet{shannon_craft_20220726} \\ 
20121102  &  557.0  &  0.1927  &  188.0  &  -3.045  &  0.5783 & \citet{petroff_frbcat_20200812} \\ 
20221106A  &  343.8  &  0.2044  &  35.0  &  -3.0756  &  -0.4463 & \citet{shannon_craft_20221113} \\ 
20240215A  &  549.5  &  0.21  &  42.8  &  -2.8292  &  1.2258  & \citet{law_dsa_20240601} \\ 
20210117A  &  728.95  &  0.214  &  34.4  &  -2.746  &  -0.2819 & \citet{shannon_craft_20210119} \\ 
20221027A  &  452.5  &  0.229  &  40.59  &  -2.9893  &  1.2584 & \citet{law_dsa_20230810} \\ 
20191001  &  507.9  &  0.234  &  44.7  &  -2.7654  &  -0.9555 & \citet{shannon_2019_ASKAP191001} \\ 
20190714  &  504.13  &  0.2365  &  39.0  &  -2.9275  &  -0.2273 & \citet{shannon_2019_ASKAP190714} \\ 
20221101B  &  490.7  &  0.2395  &  192.35  &  -2.7434  &  1.2336 & \citet{law_dsa_20230810}\\ 
20190520B  &  1210.3  &  0.241  &  113.0  &  -2.8617  &  -0.197 & \citet{niu_transient_20210925} \\ 
20220825A  &  651.24  &  0.241397  &  79.7  &  -2.7786  &  1.2668 & \citet{sherman_2024_deep} \\ 
20191228  &  297.5  &  0.2432  &  33.0  &  -2.7408  &  -0.5165 & \citet{petroff_frbcat_20200812} \\ 
20220307B  &  499.27  &  0.248123  &  135.7  &  -2.7333  &  1.26 & \citet{sherman_2024_deep} \\ 
20221113A  &  411.4  &  0.2505  &  115.37  &  -3.0585  &  1.2271 & \citet{law_dsa_20240601} \\ 
20220831A  &  1146.25  &  0.262  &  187.94  &  -2.7475  &  1.2258 & \citet{law_dsa_20230810} \\ 
20231123B  &  396.7  &  0.2625  &  33.81  &  -2.8594  &  1.2354 & \citet{law_dsa_20240601} \\ 
20230307A  &  608.9  &  0.271  &  29.47  &  -2.9347  &  1.2513 & \citet{law_dsa_20240601} \\ 
20221116A  &  640.6  &  0.2764  &  196.17  &  -3.1169  &  1.2681 & \citet{law_dsa_20240601} \\ 
20220105A  &  580.0  &  0.2785  &  22.0  &  -2.8986  &  0.3921 & \citet{shannon_craft_20220126} \\ 
20210320C  &  384.8  &  0.2797  &  42.0  &  -2.9037  &  -0.2814 & \citet{shannon_craft_20230204} \\ 
20221012A  &  441.08  &  0.284669  &  54.4  &  -2.8149  &  1.2309 & \citet{sherman_2024_deep} \\ 
20240229A  &  491.15  &  0.287  &  29.52  &  -2.9438  &  1.2335 & \citet{law_dsa_20240402} \\ 
20190102  &  364.5  &  0.2913  &  57.3  &  -2.7664  &  -1.3871 & \citet{macquart_census_2020} \\ 
20220506D  &  396.97  &  0.30039  &  89.1  &  -2.7715  &  1.2711 & \citet{sherman_2024_deep} \\ 
20230501A  &  532.5  &  0.301  &  180.18  &  -2.746  &  1.2378 & \citet{law_dsa_20240601} \\ 
20180924  &  361.42  &  0.3214  &  40.5  &  -2.7622  &  -0.7138 & \citet{bannister_single_2019} \\ 
20230626A  &  451.2  &  0.327  &  32.51  &  -2.8674  &  1.2415 & \citet{law_dsa_20240601} \\ 
20180301  &  536.0  &  0.3304  &  152.0  &  -3.0331  &  0.0815 & \citet{petroff_frbcat_20200812} \\ 
20231220A  &  491.2  &  0.3355  &  44.55  &  -2.9974  &  1.2856 & \citet{law_dsa_20240601} \\ 
20211203C  &  635.0  &  0.3439  &  63.0  &  -2.9036  &  -0.5477 & \citet{shannon_craft_20230204} \\ 
20220208A  &  437.0  &  0.351  &  128.79  &  -2.7663  &  1.2224 & \citet{law_dsa_20230810} \\ 
20220726A  &  686.55  &  0.361  &  111.42  &  -3.0556  &  1.2205 & \citet{law_dsa_20230810} \\ 
20230902A  &  440.1  &  0.3619  &  34.0  &  -3.0809  &  -0.8261 & \citet{shannon_craft_20230902} \\ 
20200906  &  577.8  &  0.3688  &  36.0  &  -3.0881  &  -0.2458 & \citet{bhandari_characterizing_2022} \\ 
20240119A  &  483.1  &  0.37  &  30.98  &  -2.8804  &  1.2499 & \citet{law_dsa_20240601} \\ 
20220330D  &  468.1  &  0.3714  &  57.83  &  -2.9511  &  1.2279 & \citet{law_dsa_20230810} \\ 
20190611  &  321.4  &  0.378  &  43.67  &  -2.7684  &  -1.3857 & \citet{macquart_census_2020} \\ 
20220501C  &  449.5  &  0.381  &  31.0  &  -2.7316  &  -0.5671 & \citet{law_dsa_20230810} \\ 
20220204A  &  612.2  &  0.4  &  46.02  &  -2.8225  &  1.2169 & \citet{law_dsa_20230810} \\ 
20230712A  &  586.96  &  0.4525  &  30.93  &  -2.9469  &  1.2664 & \citet{law_dsa_20240601} \\ 
20181112  &  589.27  &  0.4755  &  102.0  &  -2.7607  &  -0.9245 & \citet{prochaska_low_2019} \\ 
20220310F  &  462.24  &  0.477958  &  45.4  &  -2.9848  &  1.2827 & \citet{sherman_2024_deep} \\ 
20220918A  &  656.8  &  0.491  &  41.0  &  -3.1211  &  -1.2359 & \citet{shannon_craft_20220919} \\ 
20190711  &  593.1  &  0.522  &  56.4  &  -2.1349  &  -1.4025 & \citet{petroff_frbcat_20200812} \\ 
20230216A  &  828.0  &  0.531  &  27.05  &  -2.9595  &  0.06 & \citet{law_dsa_20240601} \\ 
20230814A  &  696.4  &  0.5535  &  137.83  &  -2.7507  &  1.2745 & \citet{ravi_dsa_20230816}\\ 
20221219A  &  706.7  &  0.554  &  38.59  &  -2.8418  &  1.2501 & \citet{law_dsa_20240601} \\ 
20190614  &  959.2  &  0.6  &  83.5  &  -3.0659  &  1.2864 & \citet{zwaniga_chimefrb_discovery_2021} \\ 
20220418A  &  623.25  &  0.622  &  37.6  &  -2.8867  &  1.2234 & \citet{sherman_2024_deep} \\ 
20190523  &  760.8  &  0.66  &  37.0  &  -2.9007  &  1.2648 & \citet{ravi_fast_2019} \\ 
20240123A  &  1462.0  &  0.968  &  113.01  &  -3.0622  &  1.2557 & \citet{law_dsa_20240601} \\ 
20221029A  &  1391.05  &  0.975  &  36.4  &  -2.9764  &  1.2645 & \citet{law_dsa_20230810} \\ 
20220610A  &  1457.624  &  1.016  &  31.0  &  -2.7331  &  -0.5849 & \citet{shannon_craft_20220612} \\ 
20230521B  &  1342.9  &  1.354  &  209.66  &  -2.7331  &  1.2416 & \citet{law_dsa_20240601} \\ 
\end{longtable}

\end{document}